\documentclass[lettersize,journal]{IEEEtran}
\usepackage{amsmath,amsfonts}
\usepackage{algorithmic}
\usepackage{algorithm}
\usepackage{array}
\usepackage[caption=false,font=normalsize,labelfont=sf,textfont=sf]{subfig}
\usepackage{textcomp}
\usepackage{stfloats}
\usepackage{url}
\usepackage{verbatim}
\usepackage{graphicx}
\usepackage{cite}
\usepackage{xcolor}

\newtheorem{theorem}{Theorem}
\newtheorem{lemma}{Lemma}
\newtheorem{proposition}{Proposition}

\newtheorem{definition}{Definition}

\usepackage{balance}

\newcommand*{\Scale}[2][4]{\scalebox{#1}{$#2$}}%

\begin{document}

\title{Human-in-the-loop Learning for Dynamic Congestion Games}

\author{Hongbo Li, and Lingjie Duan,~\IEEEmembership{Senior Member,~IEEE}
\thanks{Hongbo Li and Lingjie Duan are with the Pillar of Engineering Systems and Design, Singapore University of Technology and Design, Singapore 487372 (e-mail: hongbo\_li@mymail.sutd.edu.sg; lingjie\_duan@sutd.edu.sg).}
\thanks{This work was supported in part by the Ministry of Education, Singapore,
under its Academic Research Fund Tier 2 Grant with Award no. MOET2EP20121-
0001; in part by SUTD Kickstarter Initiative (SKI) Grant with
no. SKI 2021\_04\_07; and in part by the Joint SMU-SUTD Grant with no.
22-LKCSB-SMU-053. Part of this work was published in IEEE ISIT 2024 Conference \cite{li2024distributed}.} 
}

\markboth{IEEE Transactions on Mobile Computing}%
{Shell \MakeLowercase{\textit{et al.}}: A Sample Article Using IEEEtran.cls for IEEE Journals}

\IEEEpubid{0000--0000/00\$00.00~\copyright~2021 IEEE}

\maketitle

\begin{abstract}
Today mobile users learn and share their traffic observations via crowdsourcing platforms (e.g., Google Maps and Waze). Yet such platforms simply cater to selfish users' myopic interests to recommend the shortest path, and do not encourage enough users to travel and learn other paths for future others. Prior studies focus on one-shot congestion games without considering users' information learning, while our work studies how users learn and alter traffic conditions on stochastic paths in a human-in-the-loop manner. In a typical parallel routing network with one deterministic path and multiple stochastic paths, our analysis shows that the myopic routing policy (used by Google Maps and Waze) leads to severe under-exploration of stochastic paths. This results in a price of anarchy (PoA) greater than $2$, as compared to the socially optimal policy achieved through optimal exploration-exploitation tradeoff in minimizing the long-term social cost. Besides, the myopic policy fails to ensure the correct learning convergence about users' traffic hazard beliefs. To address this, we focus on informational (non-monetary) mechanisms as they are easier to implement than pricing. We first show that existing information-hiding mechanisms and deterministic path-recommendation mechanisms in Bayesian persuasion literature do not work with even \(\text{PoA}=\infty\). Accordingly, we propose a new combined hiding and probabilistic recommendation (CHAR) mechanism to hide all information from a selected user group and provide state-dependent probabilistic recommendations to the other user group. Our CHAR mechanism successfully ensures PoA less than \(\frac{5}{4}\), which cannot be further reduced by any other informational (non-monetary) mechanism. Besides the parallel network, we further extend our analysis and CHAR mechanism to more general linear path graphs with multiple intermediate nodes, and we prove that the PoA results remain unchanged. Additionally, we carry out experiments with real-world datasets to further extend our routing graphs and verify the close-to-optimal performance of our CHAR mechanism.
\end{abstract}


\section{Introduction}
\label{Section1}
In today's traffic networks, a random number of users arrive sequentially to decide their routing paths (e.g., the shortest path) based on real-time traffic conditions \cite{zhang2022review}. To assist new user arrivals in their decision-making process, emerging crowdsourcing applications (e.g., Google Maps and Waze) serve as important platforms to leverage users for information learning and sharing (\!\!\cite{waze,application}). These platforms aggregate the latest traffic information of stochastic paths from previous users who have recently traveled there and reported their observations, then share this information with new users. However, these platforms always myopically recommend the currently shortest path, and do not encourage enough selfish users to travel and learn other longer paths of varying traffic conditions for future others. 
Prior congestion game literature assumes a static scenario of one-shot routing by a group of users, without considering their long-term information learning in a long-term dynamic scenario (e.g., \cite{zhu2022information,vasserman2015implementing}).

To efficiently learn and share the time-varying information, multi-armed bandit (MAB) problems are developed to study the optimal exploration-exploitation among stochastic arms/paths (e.g., \cite{liu2012learning,slivkins2019introduction,gupta2021multi,bozorgchenani2021computation,wang2022truthful}). For example, \cite{bozorgchenani2021computation} utilizes MAB techniques to predict congestion in a fast-varying transportation environment. \cite{wang2022truthful} applies the MAB exploration-exploitation strategy to balance the recruitment of previously well-behaved users or new uncertain users for completing crowdsourcing tasks. Recently, some studies have extended traditional MAB problems to distributed scenarios with multiple user arrivals at the same time (e.g., \cite{li2020multi,yang2021cooperative,shi2021federated,yang2022distributed,zhu2023distributed}). For instance, \cite{yang2021cooperative} study how users make their own optimal decisions at each time slot based on the private arm information that they learned beforehand. \cite{shi2021federated} explores cooperative information exchange among all users to facilitate local model learning.
Subsequently, both \cite{yang2022distributed} and \cite{zhu2023distributed} examine the performance under partially connected communication graphs, where users communicate only with their neighbors to make locally optimal decisions. In general, these MAB solutions assume users' cooperation to align with the social planner, without possible selfish deviations to improve their own benefits.

As selfish users may not listen to the social planner, some recent congestion game works focus on Bayesian persuasion and information design to incentivize users' exploration of different paths \cite{das2017reducing,kamenica2019bayesian,mansour2022bayesian,babichenko2022regret,gollapudi2023online}. For example, \cite{mansour2022bayesian} assumes that the social planner has the complete system information and properly discloses system information to regulate users and control efficiency loss. Without perfect feedback, \cite{babichenko2022regret} estimates users' expected utilities of traveling each path based on all possible cases to improve the robustness of Bayesian persuasion. Similarly, \cite{gollapudi2023online} allows the system to learn stochastic travel latency based on users' travel data while only regulating the immediate congestion in the current time slot. In summary, all these works assume that users' routing decisions do not internally alter the long-term traffic congestion condition for future users, and only focus on exogenous information to learn dynamically.  

\IEEEpubidadjcol

It is practical to consider endogenous information variation in dynamic congestion games, where more users choosing the same path not only improve learning accuracy there (\!\!\cite{yang2009discretization}), but also produce more congestion for followers (\!\!\cite{meunier2010equilibrium,carmona2020pure}). For this new problem, we need to overcome two technical issues. The first is how to \emph{dynamically allocate users to reach the best trade-off between learning accuracy and resultant congestion costs of stochastic paths}. To minimize the long-run social travel cost, it is critical to dynamically balance multiple user arrivals' positive information learning effect and negative congestion effect to avoid both under- and over-exploration of stochastic paths.

Even provided with the socially optimal policy, we still need to ensure selfish users follow it. Thus, the second issue is how to \emph{design an informational mechanism to properly regulate users' myopic routing}. In practice, informational mechanisms (e.g., Bayesian persuasion \cite{das2017reducing,kamenica2019bayesian,mansour2022bayesian}) are non-monetary and easier to implement compared to pricing or side-payment mechanisms (e.g., \cite{zheng2020designing,Ferguson2022effective,li2024incentivizing}). There are two commonly used informational mechanisms to optimize long-run information learning in congestion games: information hiding mechanism \cite{tavafoghi2017informational,wang2020efficient,farhadi2022dynamic} and deterministic path-recommendation \cite{li2019recommending,wu2019learning,li2023congestion}. For example, \cite{tavafoghi2017informational} analyzes how users infer the expected travel cost of each path to decide routing if the system hides all traffic information. \cite{li2023congestion} assumes a publicly known process for varying each path's traffic condition, and selectively provides deterministic path recommendations to regulate users' myopic routing. We will show later that both mechanisms do not work in improving our system performance, inspiring our new mechanism design.

Our main contributions and the key novelty of this paper are summarized as follows. 
\begin{itemize}
    \item \emph{Human-in-the-loop learning for dynamic congestion games:} Prior studies focus on one-shot congestion games without information learning, while we study how users learn and alter traffic conditions on stochastic paths in a human-in-the-loop manner. We consider a typical multi-path parallel network to regulate selfish user arrivals' information learning for dynamic congestion games. When such congestion games meet human-in-the-loop learning, more users' routing on stochastic paths generates both positive learning benefits and negative congestion effects there for followers. We utilize users’ positive learning effect to negate negative congestion, which fundamentally extends traditional one-shot congestion games (e.g., \cite{mansour2022bayesian,gollapudi2023online} and \cite{wu2019learning}).
    \item \emph{Policies comparison via PoA analysis:} 
    To minimize the social cost, we formulate the optimization problems for both myopic and socially optimal policies as Markov decision processes (MDP). Our analysis shows that the myopic routing policy (used by Google Maps and Waze) leads to severe under-exploration of stochastic paths. This results in a price of anarchy (PoA) greater than $2$, as compared to the socially optimal policy achieved through optimal exploration-exploitation tradeoff in minimizing the long-term social cost. 
    \item \emph{Learning convergence and new mechanism design:} We first prove that the socially optimal policy ensures correct convergence of users’ traffic hazard beliefs on stochastic paths, while the myopic policy cannot. Then we show that existing information-hiding and deterministic-recommendation mechanisms in Bayesian persuasion literature make ${\text{PoA}=\infty}$. Accordingly, we propose a combined hiding and probabilistic recommendation (CHAR) mechanism to hide all information from a selected user group and provide state-dependent probabilistic recommendations to the other user group. Our CHAR mechanism successfully ensures PoA less than \(\frac{5}{4}\), which cannot be further reduced by any other informational mechanism.
    \item \emph{Extensions to linear path graph:} Besides the basic parallel transportation network between two nodes, we further extend our analysis and CHAR mechanism to encompass any linear path graph with multiple intermediate nodes. In this extended model, myopic users also consider their future travel costs when making current routing decisions, which adds complexity to their decision analysis. Nonetheless, we successfully prove that the PoA caused by users' myopic routing policy remains greater than $2$, and our CHAR mechanism continues to effectively reduce PoA to less than $\frac{5}{4}$. Finally, we conduct experiments using real-world datasets to show the close-to-optimal average-case performance of our CHAR mechanism.
\end{itemize}

The rest of the paper is organized as follows. In Section~\ref{section2}, we formulate the dynamic congestion model and the human-in-the-loop learning model for the system. Then we formulate the optimization problems for both myopic policy and socially optimal policy in Section~\ref{section3}. After that, in Section~\ref{section4} we analytically compare the two policies via PoA analysis. Based on the analysis, we propose our CHAR mechanism in Section~\ref{section5}. Subsequently, we extend the system model and analysis to a general linear path graph in Section~\ref{section6} and verify the performance of our CHAR mechanism in Section~\ref{section7}. Finally, Section~\ref{section8} concludes the paper. For ease of reading, we summarize all the key notations in Table \ref{notation_table}.

\section{System Model}\label{section2}

\begin{figure}[t]
    \centering
    \includegraphics[width=0.36\textwidth]{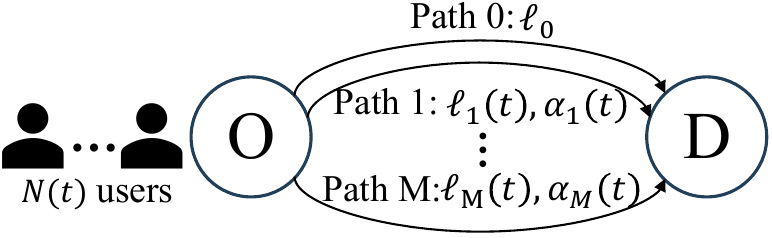}
    \captionsetup{font={footnotesize}}
    \caption{Dynamic congestion model: within each time slot $t\in\{0,1,\cdots\}$, a random number $N(t)$ of users arrive at origin O to decide among safe path~0 and any stochastic path ${i\in\mathcal{M}:=\{1,\cdots, M\}}$ to travel to destination D.}
    \label{fig:congestion_game}
\end{figure}

In this section, we first introduce the dynamic congestion model for a typical parallel multi-path network between two nodes O and D as in existing congestion game literature (e.g., \cite{tavafoghi2017informational, wu2019learning} and \cite{li2023congestion}). Then we introduce the human-in-the-loop learning model for the crowdsourcing platform. Note that we have extended our congestion model and key results to any general linear path graphs later in Section \ref{section6}.
As shown in Fig.~\ref{fig:congestion_game}, we consider an infinite discrete time horizon ${t\in\{0,1,\cdots\}}$ to model the dynamic congestion game. Within each time slot $t$, a stochastic number $N(t)\in\ [N_{min},N_{max}]$ of atomic users, where $\mathbb{E}[N(t)]=\Bar{N}$, arrive at origin O to decide their paths among $M+1$ paths to travel to destination~D.

\begin{table}[!t]
\renewcommand{\arraystretch}{1.3}
\caption{Key notations and their meanings in the paper}
\label{notation_table}
\centering
\begin{tabular}{|c|m{0.33\textwidth}|}
\hline
\textbf{Notation} & \textbf{Meaning}\\
\hline
\hline
$N(t)$ & The number of user arrivals at time $t$.\\
\hline
$N_{min},N_{max}$ & The lower and upper bounds of $N(t)$.\\
\hline
$\Bar{N}$ & The expectation value of $N(t)$.\\
\hline
$\mathcal{M}$ & The set of stochastic paths.\\
\hline
$\ell_0$ & The fixed travel latency of path $0$.\\
\hline
$\ell_i(t)$ & The travel latency of path $i$ at time $t$.\\
\hline
$n_i(t)$ & The number of users traveling path $i$ at time $t$.\\
\hline
$\alpha_i(t)$ & The correlation coefficient of stochastic path $i$.\\
\hline
$\alpha_H,\alpha_L$ & The high and low hazard states on stochastic paths.\\
\hline
$\Bar{x}$ & The long-run expected probability of $\alpha_i(t)=\alpha_H$.\\
\hline
$\mathbb{P}(\Bar{x})$ & The distribution of $\Bar{x}$.\\
\hline
$y_i(t)$ & Users' observation summary of path $i$ at time $t$.\\
\hline
$q_{H}(\cdot),q_{L}(\cdot)$ & The probabilities for $n_i(t)$ users to report $y_i(t)=1$ under $\alpha_H$ and $\alpha_L$, respectively. \\
\hline
$x_i(t),x_i'(t)$ & The prior and posterior hazard beliefs of stochastic path $i$ at time $t$.\\
\hline
$\mathbf{L}(t)$ & The expected travel latency set of all $M+1$ paths.\\
\hline
$\mathbf{x}(t)$ & The hazard belief set of $M$ stochastic paths.\\
\hline
$c_i(\cdot)$ & Each user's travel cost on path $i$ at time $t$.\\
\hline
$V(\cdot)$ & The variance cost at time $t$.\\
\hline
$C(\cdot)$ & The long-term cost function.\\
\hline
$\rho$ & Discount factor.\\
\hline
$\pi(t)$ & The private path recommendation for each user under our CHAR mechanism at time $t$.\\
\hline
\end{tabular}
\end{table}

\subsection{Dynamic Congestion Model}

Following the traditional routing game literature (\!\!\cite{kremer2014implementing,das2017reducing,li2023congestion}), we model a safe path 0 with a fixed travel latency, which is denoted by $\ell_0$. While the other paths are stochastic with time-varying travel latencies, denoted by $\ell_i(t)$ at time $t\in\{1,2,\cdots\}$, where $i\in\mathcal{M}:=\{1,\cdots,M\}$. 
According to prior research on travel delay estimation (e.g., \cite{ban2009delay,alam2019prediction}), the next $\ell_i(t+1)$ at $t+1$ is correlated with both current $\ell_i(t)$ and the number of atomic users traveling this path, which is denoted by $n_i(t)$. For example, more users traveling on the same path results in slower driving speeds and longer queueing times at traffic lights, thereby increasing the latency for later users. Define this general correlation function as:
\begin{align}\label{ell_2}
    \ell_i(t+1)=f\big(\ell_i(t),n_i(t),\alpha_i(t)\big),
\end{align}
where $\alpha_i(t)$ is the correlation coefficient, measuring the leftover flow to be served over time. Note that correlation function $f(\cdot)$ is defined as a general increasing function in $\ell_i(t),n_i(t)$ and $\alpha_i(t)$.

As in existing congestion game literature (\!\!\cite{wu2019learning,farhadi2022dynamic,li2023congestion}), we model $\alpha_i(t)$ to follow a memoryless stochastic process (not necessarily a time-invariant Markov chain) to alter as below:
\begin{align*}
    \alpha_i(t)=\begin{cases}
    \alpha_H,\text{if high hazard (bad) condition on path $i$ at }t,\\
    \alpha_L,\text{if low hazard (good) condition on path $i$ at }t,
    \end{cases}
\end{align*}
where ${\alpha_H\in[1,+\infty)}$ and ${\alpha_L\in[0,1)}$.
Unlike these works assuming that the social planner knows the probability distribution of $\alpha_i(t)$ beforehand, we relax this assumption to keep it unknown in our work. For ease of exposition, we only consider two states here, while our later analysis and mechanism design can be extended to accommodate multiple states of $\alpha_i(t)$.

We denote the expected steady state of correlation coefficient $\alpha_i(t)$ in the long run by
\begin{align}
    \mathbb{E}[\alpha_i(t)|\Bar{x}]=\Bar{x}\cdot \alpha_H+(1-\Bar{x})\cdot \alpha_L, \label{bar_alpha}
\end{align}
where $\Bar{x}$ is the long-run expected probability of $\alpha_i(t)=\alpha_H$ yet the exact value of $\Bar{x}$ is unknown to users. They can only know $\Bar{x}$ to satisfy a distribution $\mathbb{P}(\Bar{x})$. Here we assume $\mathbb{P}(\Bar{x})$ is not extreme, such that users may consider traveling on any stochastic path $i$ under $\Bar{x}\sim\mathbb{P}(\Bar{x})$. To accurately estimate $\mathbb{E}[\alpha_i(t)|\Bar{x}]$, the system expects users to learn the real steady state $\Bar{x}$ by observing real-time traffic conditions.

\subsection{Human-in-the-loop Learning Model}
When traveling on stochastic path $i$, users cannot observe $\alpha_i(t)$ directly but a traffic hazard event (e.g., ‘black ice’ segments or jamming). At the same time, \cite{venanzi2013crowdsourcing} shows that each user has a noisy observation of the hazard. Hence, our system uses a majority vote to fuse all their observation reports of stochastic path $i$ into a hazard summary ${y_i(t)\in\{0,1,\emptyset\}}$ during time $t$. Specifically,
\begin{itemize}
    \item ${y_i(t) = 1}$ tells that most of the current ${n_i(t)}$ users observe a traffic hazard on stochastic path $i$ at time $t$.
    \item ${y_i(t)=0}$ tells that most of the current ${n_i(t)}$ users observe no traffic hazard on stochastic path $i$ at time $t$.
    \item ${y_i(t)=\emptyset}$ tells that there is no user traveling on path $i$ (i.e., ${n_i(t)=0}$), without any new observation. 
\end{itemize}

However, the fused summary $y_i(t)$ by $n_i(t)$ users can still be inaccurate. Given $n_i(t)$, we define the group probabilities under $\alpha_i(t)=\alpha_H$ and $\alpha_i(t)=\alpha_L$ for observing a hazard below:
\begin{align}
    q_H(n_i(t))&=\mathbf{Pr}\big(y_i(t)=1|\alpha_i(t)=\alpha_H, n_i(t)\big),\label{F1F0}\\
    q_L(n_i(t))&=\mathbf{Pr}\big(y_i(t)=1|\alpha_i(t)=\alpha_L, n_i(t)\big).\notag
\end{align}
According to \cite{yang2009discretization}, ${q_H(n_i(t))\in[0,1]}$ increases with $n_i(t)$, because more users help spot hazard ${\alpha_i(t)=\alpha_H}$. Similarly, ${q_L(n_i(t))}\in[0,1]$ decreases with $n_i(t)$ under ${\alpha_i(t)=\alpha_L}$.

We summarize the current routing decisions and observations of all paths as vectors $\mathbf{n}(t)=\{n_i(t)|i\in\{0,\cdots, M\}\}$ and ${\mathbf{y}(t)=\{y_i(t)|i\in\mathcal{M}\}}$, respectively. To guide trip advisory at time $t$, the crowdsourcing platform needs to memorize all the historical data of ${(\mathbf{n}(1),\cdots,\mathbf{n}(t-1))}$ and ${(\mathbf{y}(1),\cdots,\mathbf{y}(t-1))}$, which keep growing over time. Following the memoryless property \cite{li2019recommending}, we use Bayesian inference to equivalently summarize these data into a prior hazard belief ${x_i(t)\in[0,1]}$ for any stochastic path~$i$:
\begin{align}
    {x_i(t)=\mathbf{Pr}\big(\alpha_i(t)=\alpha_H|x_i(t-1),y_i(t-1),n_i(t-1)\big)},\label{x(t)}
\end{align}
which indicates the probability of ${\alpha_i(t)=\alpha_H}$ at time $t$.

Next, we explain how our human-in-the-loop learning model works in the time horizon.

\begin{itemize}
    \item At the beginning of each time slot $t$, the crowdsourcing platform publishes latest latency $\mathbb{E}[\ell_i(t)|x_i(t-1),y_i(t-1)]$ and prior hazard belief $x_i(t)$ in (\ref{x(t)}) on any stochastic path $i$. 
    \item During time $t$, ${N(t)}$ users arrive to decide $n_i(t)$ on each stochastic path~$i$ and travel there to return their observation summary $y_i(t)$. Based on $y_i(t)$, the platform next updates prior belief $x_i(t)$ to a posterior belief $x_i'(t)=\mathbf{Pr}(\alpha_i(t)=\alpha_H|x_i(t),y_i(t),n_i(t))$. For example, if ${y_i(t)=1}$, we have
    \begin{align}       
        {x_i'(t)}
        &{=\text{Pr}(\alpha_i(t)=\alpha_H|x_i(t),n_i(t),y_i(t)=1)}\notag\\
    &=\frac{\Scale[0.94]{\text{Pr}(y_i(t)=1|\alpha_i(t)=\alpha_H)\text{Pr}(\alpha_i(t)=\alpha_H)}}{\Scale[0.94]{\sum_{j\in\{\alpha_H,\alpha_L\}}\text{Pr}(y_i(t)=1|\alpha_i(t)=j)\text{Pr}(\alpha_i(t)=j)}}\notag\\
    &=\frac{{x_i(t)q_H(n_i(t))}}{{x_i(t)q_H(n_i(t))+(1-x_i(t))q_L(n_i(t))}}\label{x'(t)}
    \end{align}
    where the second equation is due to Bayes' Theorem and Law of total probability.
    Similarly, if $y_i(t)=0$, we obtain
    \begin{align*}       
        x'(t)=\frac{\Scale[0.95]{ x(t)\big(1-q_H(n(t))\big)}}{\Scale[0.95]{ x(t)\big(1-q_H(n(t))\big)+(1-x(t))\big(1-q_L(n(t))}\big)}.
    \end{align*}
    Note that we keep $x'(t)=x(t)$ if $y_i(t)=\emptyset$, as there is no observation to update $x'(t)$.
    \item At the end of time $t$, the platform updates the expected correlation coefficient given posterior belief $x'_i(t)$:
    \begin{align}
        {\mathbb{E}[\alpha_i(t)|x_i'(t)]=x_i'(t)\alpha_H+(1-x_i'(t))\alpha_L.}\label{E_alpha}
    \end{align}
    Combing (\ref{E_alpha}) with (\ref{ell_2}), we obtain expected travel latency
    \begin{align}
        \Scale[0.96]{
        \mathbb{E}[\ell_i(t+1)|x_i'(t)]=\mathbb{E}\big[f\big(\mathbb{E}[\ell_i(t)|x'_i(t-1)],n_i(t),\alpha_i(t)\big)\big]}\label{E_ell_2}
    \end{align}
     on stochastic path $i$ for next time slot ${t+1}$.
\end{itemize}
    Finally, the platform updates ${x_i(t+1)=x_i'(t)}$ and repeats  above process in $t + 1$.

\section{Problem Formulations for Myopic and Socially Optimal Policies}\label{section3}

Based on the dynamic congestion and human-in-the-loop learning models in Section \ref{section2}, we next formulate the optimization problems for both myopic and socially optimal policies, respectively. Note that the myopic policy is widely used in existing crowdsourcing platforms (e.g., Google Maps and Waze), which is defined as follows.
\begin{definition}[Myopic Policy]
Under the myopic policy, the platform recommends a path to each user to minimize their current travel cost.
\end{definition}

While the socially optimal policy minimizes the long-term social travel cost for all users.

\subsection{Problem Formulation for the Myopic Policy}
In this subsection, we focus on the myopic policy used by Google Maps and Waze, which post all the collected useful information to the public. First, we summarize all stochastic paths' expected latencies and hazard beliefs into vectors
\begin{align*}
    \mathbf{L}(t)&=\{\mathbb{E}[\ell_i(t)|x_i'(t-1)]|i\in\mathcal{M}\},\\
    \mathbf{x}(t)&=\{x_i(t)|i\in\mathcal{M}\},
\end{align*}
respectively. Let $n_i^{(m)}(t)\in\{0,\cdots, N(t)\}$ define the (exploration) number of users traveling on any path~${i\in\{0,\cdots,M\}}$ allocated by the myopic policy from $N(t)$ users, where the superscript $(m)$ means the myopic policy. We use ${\mathbf{n}^{(m)}(t)}$ to summarize all the ${n_i^{(m)}(t)}$ under the myopic policy.

Following the typical travel cost model as in congestion game literature \cite{das2017reducing,tavafoghi2017informational}, for each of the $n_0(t)$ users on safe path 0, his expected travel cost consists of travel latency $\ell_0$ and congestion cost ${n_0(t)}$ caused by other users on this path:
\begin{align}
    {c_0(n_0(t))=\ell_0+n_0(t).}\label{cm_1}
\end{align}

For each user on stochastic path $i\in\mathcal{M}$, besides the expected travel latency ${\mathbb{E}[\ell_i(t)|x_i'(t-1)]}$ caused by past users and the immediate congestion cost ${n_i(t)}$ in traditional congestion games without learning (\!\!\cite{das2017reducing,tavafoghi2017informational}), he also faces an extra variance cost ${V(n_i(t-1))}$ due to former ${n_i(t-1)}$ users' imperfect observation summary $y_i(t-1)$ (\!\!\cite{smith2018bayesian,wu2019learning}). $V(n_i(t-1))$ tells how much the gap (between the ${n_i(t-1)}$ reports fused result and the reality) adds to the total cost by misleading decision making of future users. Then his travel cost on stochastic path $i$ is:
\begin{align}
    {c_i(n_i(t))=\label{cm_2}\mathbb{E}[\ell_i(t)|x_i'(t-1)]+n_i(t)+V\big(n_i(t-1)\big),}
\end{align}
where ${V(n_i(t-1))}$ is a general decreasing function of ${n_i(t-1)}$, as more users improve the learning accuracy on stochastic path $i$.

Based on each user's individual cost (\ref{cm_1}) of safe path 0 and (\ref{cm_2}) of stochastic path $i$, we next analyze myopic policy $n_i^{(m)}(t)$. For ease of exposition, we first consider $M=1$ in a basic two-path network to solve and explain ${n_1^{(m)}(t)}$ below. 
\begin{lemma}\label{lemma:n1m}
Under the myopic policy, given expected travel latency ${\mathbb{E}[\ell_1(t)|x_1'(t-1)]}$ and hazard belief ${x_1(t)}$ of stochastic path 1, the exploration number is:
\begin{align}
    {n^{(m)}_1(t)=}\begin{cases}
        {0}, \quad\quad\quad\quad\ \ \quad\text{if }{c_1(0)\geq c_0(N(t))},\\
        {N(t)}, \quad\quad\quad\quad\text{if }{c_1(N(t))\leq c_0(0)},\\
        \frac{{N(t)}}{{2}}+\frac{{c_0(0)- c_1(0)}}{{2}},\quad\text{ otherwise}.
    \end{cases}\label{nm(t)}
\end{align}
\end{lemma}

The proof of Lemma 1 is given in Appendix A of the supplement material.
If the minimum expected travel cost of stochastic path~1 is larger than the maximum cost of path~0, all the $N(t)$ users will choose path~0 with ${n_1^{(m)}(t)=0}$ in the first case of (\ref{nm(t)}). Otherwise, in the last two cases of (\ref{nm(t)}), there is always a positive ${n_1^{(m)}(t)\geq 1}$ number of users traveling on stochastic path 1 to learn and update ${x_1(t+1)}$ and ${\mathbb{E}[\ell_1(t+1)|x_1'(t)]}$ there. 
When there are $M \geq 2$ stochastic paths, $N(t)$ user arrivals myopically decide on the exploration number $n_i(t)$ for each path $i$, as described in (\ref{nm(t)}), in an attempt to balance the expected cost $c_i(n_i^{(m)}(t))=c_j(n^{(m)}_j(t))$ for any path $j \neq i$ at the equilibrium. Under the constraint $\sum_{i=0}^Mn_i^{(m)}(t)=N(t)$, this system of $M+1$ linear equations in $M+1$ unknowns can be solved by Gaussian elimination.

Based on the above analysis, we further examine the long-run social cost of all users since the current time $t$. For current ${N(t)}$ user arrivals, their immediate social cost under the myopic policy is 
\begin{align}
    {c(\mathbf{n}^{(m)}(t))=\sum_{i=0}^Mn_i^{(m)}(t)c_i(n_i^{(m)}(t))}\label{cm}
\end{align}
with individual cost $c_i(\cdot)$ in (\ref{cm_1}) or (\ref{cm_2}).
Define the long-term $\rho$-discounted social cost function under the myopic policy to be $C^{(m)}(\mathbf{L}(t),\mathbf{x}(t),N(t))$. Based on (\ref{cm}), we leverage the Markov decision process (MDP) to formulate:
\begin{align}
    &{C^{(m)}\big(\mathbf{L}(t),\mathbf{x}(t),N(t)\big)=}\label{Cm}\\ &{c(\mathbf{n}^{(m)}(t))+\rho \mathbb{E}\big[C^{(m)}\big(\mathbf{L}(t+1),\mathbf{x}(t+1),N(t+1)\big)}\big],\notag
\end{align}
where the update of each travel latency ${\mathbb{E}[\ell_i(t+1)|x_i'(t)]}$ and hazard belief $x_i(t+1)$ are given in (\ref{x'(t)}) and (\ref{E_ell_2}), depending on current ${N(t)}$ users' routing decision ${n_i^{(m)}(t)}$ and the possible observation summary ${y_i(t)\in\{0,1,\emptyset\}}$ on path $i$. Here the discount factor ${\rho\in(0,1)}$ is widely used to discount future travel costs to its present cost \cite{roughgarden2005selfish}.

\subsection{Problem Formulation for Socially Optimal Policy}
Define ${n_i^*(t)\in\{0,1,\cdots,N(t)\}}$ to be the exploration number under the socially optimal policy, and let vector $\mathbf{n}^*(t)$ summarize all the $n_i^*(t)$. Different from the myopic policy that only minimizes each user's travel cost, the socially optimal policy aims to well control the overall congestion cost and information learning to minimize the long-run expected social cost function.

Define the long-run expected social cost function under the socially optimal policy to be ${C^*(\mathbf{L}(t),\mathbf{x}(t),N(t))}$.
Under the socially optimal policy, the update of $\mathbf{L}(t)$ and $\mathbf{x}(t)$ also depend on the optimal exploration number set $\mathbf{n}^*(t)$. For any stochastic path $i$,
\begin{itemize}
    \item If $n_i^*(t)=0$, i.e., the platform asks no user to choose this stochastic path~$i$, then $y_i(t)=\emptyset$ and the next hazard belief $x_i(t+1)=x_i(t)$ keeps unchanged. 
    \item Otherwise, $x_i(t+1)$ is updated based on $y_i(t)=1$ or $0$. 
\end{itemize}
After that, the expected travel latency $\mathbb{E}[\ell_i(t+1)|x_i'(t)]$ of each stochastic path $i$ is updated according to (\ref{E_ell_2}).

Based on the above analysis, we formulate the long-term objective function under the socially optimal policy as:
\begin{align}
    &C^*\big(\mathbf{L}(t),\mathbf{x}(t),N(t)\big)=\label{C*}\\ &\min_{\mathbf{n}^*(t)}\ c(\mathbf{n}^*(t))+\rho \mathbb{E}\big[C^*\big(\mathbf{L}(t+1),\mathbf{x}(t+1),N(t+1)\big)\big], \notag
\end{align}
where the immediate social cost $c(\mathbf{n}^*(t))$ is similarly defined as in (\ref{cm}).
Note that (\ref{C*}) is non-convex and incurs the curse of dimensionality under the infinite time horizon \cite{bertsimas1997introduction}. 

To derive an approximated solution $\mathbf{n}^*(t)$, we combine successive approximation \cite{ross2014introduction} and exhaustive search \cite{yao1980efficient} to formulate the following iterative algorithm, which iterates over $Q$ steps. In each step $q\in\{1,\cdots,Q\}$, we update the current solution $\mathbf{n}_q(t)$ by solving the problem $C_q(\mathbf{L}(t),\mathbf{x}(t),N(t))$. 

Initially, we define
\begin{align*}
    C_0(\mathbf{L}(t),\mathbf{x}(t),N(t))=\min_{\mathbf{n}_0(t)} \ c(\mathbf{n}_0(t)),
\end{align*}
which is a linear optimization problem and can be solved by the ellipsoid method \cite{bertsimas1997introduction}.
Then for any step $q\geq 1$, let
\begin{align*}
    &C_q(\mathbf{L}(t),\mathbf{x}(t),N(t))=\\ &\min_{\mathbf{n}_q(t)} \ c(\mathbf{n}_q(t))+\rho \mathbb{E}\big[C_{q-1}\big(\mathbf{L}(t+1),\mathbf{x}(t+1),N(t+1)\big)\big],
\end{align*}
which can be solved by the exhaustive search in the finite discrete action space \cite{yao1980efficient}. According to \cite{ross2014introduction}, $\mathbf{n}_q(t)$ will finally converge to the actual $\mathbf{n}^*(t)$ as $q$ increases.


\section{Policies Comparison via PoA Analysis}\label{section4}
In this section, we first analyze the monotonicity of the two policies with respect to hazard belief ${x_i(t)}$ and variance cost ${V(\cdot)}$. Then, we show that the myopic policy misses both exploration and exploitation over time, as compared to the socially optimal policy. Finally, we prove $\text{PoA}\geq 2$ caused by the myopic policy, which implies at least doubled total travel cost for all users and motivates our mechanism design in Section~\ref{section5}.

We first analyze the monotonicity of cost functions under both policies in (\ref{Cm}) and (\ref{C*}) below.
\begin{lemma}\label{lemma:mono_C}
The long-term cost functions (\ref{Cm}) and (\ref{C*}) under both policies increase with expected travel latency ${\mathbb{E}[\ell_i(t)|x_i'(t-1)]}$, hazard state $x_i(t)$ and users' variance cost ${V(\cdot)}$ on stochastic path $i$. 
\end{lemma}

The proof of Lemma 2 is given in Appendix B of the supplement material. As hazard belief ${x_i(t)}$ increases, the expected correlation coefficient ${\mathbb{E}[\alpha_i(t)|x_i'(t)]}$ will also increase, which incurs a larger travel cost for future users choosing stochastic path $i$. As variance cost ${V(\cdot)}$ enlarges, users at any time $t$ face a greater error to estimate the travel latency on path $i$, which incurs a larger travel cost in (\ref{cm_2}).


\begin{figure}[t]
    \centering
    \includegraphics[width=0.48\textwidth]{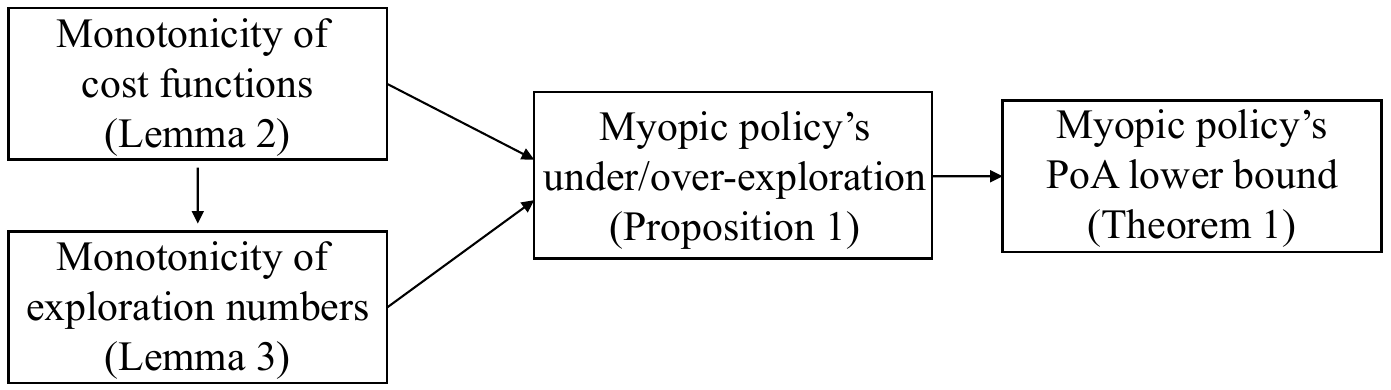}
    \caption{Theoretical results of the myopic policy and the socially optimal policy in Section~\ref{section4}.}
    \label{fig:lemmas}
\end{figure}

Based on the monotonicity of cost functions in Lemma~\ref{lemma:mono_C}, we then analyze the monotonicity of their corresponding exploration numbers $n_i^{(m)}(t)$ and $n_i^*(t)$ in the next lemma.
\begin{lemma}\label{lemma:n*-nm}
Both exploration numbers ${n_i^{(m)}(t)}$ and ${n_i^*(t)}$ decrease with ${x_i(t)}$ and ${V(\cdot)}$. While the difference $n_i^*(t)-n_i^{(m)}(t)$ increases with ${x_i(t)}$ but decreases with ${V(\cdot)}$.
\end{lemma}

The proof of Lemma 3 is given in Appendix C of the supplement material. Intuitively, as hazard belief ${x_i(t)}$ increases, the myopic policy becomes less willing to explore stochastic path $i$, which has a larger immediate travel cost than safe path 0, than the socially optimal policy. However, as ${V(\cdot)}$ enlarges, users' explorations will incur extra error costs, making both policies unwilling to explore. 
Lemma \ref{lemma:n*-nm} also tells that longer expected latency $\mathbb{E}[\ell_i(t)|x_i'(t-1)]$, stronger hazard belief $x_i(t)$, and enlarged variance cost ${V(\cdot)}$ may hinder allocating more users to explore stochastic path $i$.

\subsection{Exploration and Exploitation Comparison}
Thanks to Lemma \ref{lemma:n*-nm}, now we are ready to analytically compare $n_i^{(m)}(t)$ with $n_i^*(t)$. Before that, we first define the myopic routing policy's under/over-exploration of stochastic path $i$ as compared to the social optimum in the following.
\begin{definition}[Under/over exploration]\label{def:under-over}
The myopic policy under-explores stochastic path $i$ if $n_i^{(m)}(t)< n_i^*(t)$, and over-explores if $n_i^{(m)}(t)\geq n_i^*(t)$ at time $t$.
\end{definition}

Based on Definition \ref{def:under-over}, we next prove that the myopic policy misses both proper exploration and exploitation of stochastic path $i$, as hazard belief $x_i(t)$ dynamically changes over time.
\begin{proposition}\label{Prop:explore}
There exists a belief threshold ${x_{th}\in(0,1)}$, such that the myopic policy will over-explore stochastic path~$i$ (with ${n_i^{(m)}(t)\geq n_i^*(t)}$) if ${x_i(t)<x_{th}}$, and will under-explore stochastic path~$i$ (with ${n_i^{(m)}(t)<  n_i^*(t)}$) if ${x_i(t)\geq x_{th}}$, as compared to the socially optimal policy. This belief threshold ${x_{th}}$ increases with ${V(\cdot)}$.
\end{proposition}

The proof of Proposition 1 is given in Appendix D of the supplement material. If there is a strong hazard belief ${x_i(t)\geq x_{th}}$ of ${\alpha_i(t)=\alpha_H}$, the myopic policy is not willing to explore stochastic path $i$ due to longer latency. However, the socially optimal policy still allocates some users to learn possible ${\alpha_L}$ future others to exploit. In contrast, given weak ${x_i(t)<x_{th}}$, myopic users flock to stochastic path~$i$ without considering future congestion, while the social optimum may exploit safe path 0 to reduce congestion on path~$i$. According to Lemma \ref{lemma:n*-nm}, as error cost ${V(\cdot)}$ increases, the socially optimal policy also becomes less willing to explore stochastic path~$i$, leading to a belief threshold ${x_{th}}$.

\begin{figure}[t]
    \centering
    \includegraphics[width=0.35\textwidth]{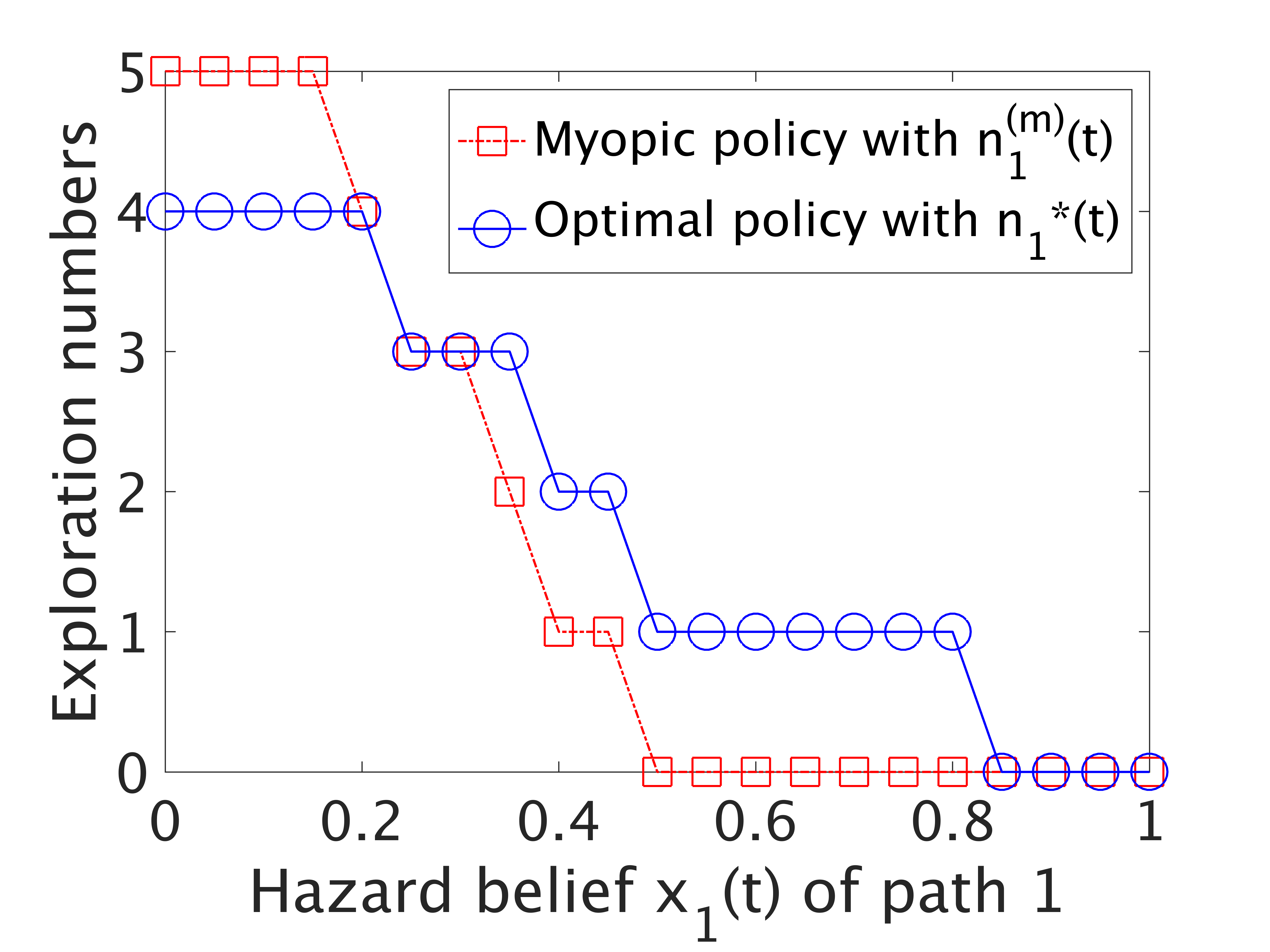}
    \caption{Exploration numbers $n_1^*(t)$ under the socially optimal policy and $n_1^{(m)}(t)$ under the myopic policy versus hazard belief $x_1(t)$ in an illustrative two-path transportation network with $M=1$.} 
    \label{fig:exploration_x}
\end{figure}

To illustrate this, in Fig. \ref{fig:exploration_x}, we simulate Fig. \ref{fig:congestion_game} using an illustrative basic two-path network with $M=1$. We plot the myopic exploration number $n_1^{(m)}(t)$ in (\ref{nm(t)}) and the optimal $n_1^*(t)$ derived by (\ref{C*}) versus hazard belief $x_1(t)$ of stochastic path~1. These two thresholds are very different as shown in Fig. \ref{fig:exploration_x}. Here we set $\ell_0=15,\Bar{N}=5,\mathbb{E}[\ell_1(t-1)]=20, \alpha_H=1.5,\alpha_L=0.05$ at current time $t$. Here $q_H(n_1(t))$ in (\ref{F1F0}) follows a normal distribution’s CDF with mean $0.3$ and variance $1$, and we properly choose error function $V(n(t-1))=\min\{\frac{10}{n(t-1)},20\}$. Given the belief threshold $x_{th}=0.3$ here, if the hazard belief $x_1(t)<x_{th}$, we have the myopic exploration threshold $n_1^{(m)}(t)$ larger than the optimal $n_1^*(t)$ to over-explore stochastic path 1. 
If $x_1(t)\geq x_{th}$, the myopic exploration threshold satisfies $n_1^{(m)}(t)$ smaller than the optimal $n_1^*(t)$ to under-explore. This result is consistent with Proposition \ref{Prop:explore}.

\subsection{PoA Analysis}
After comparing the two policies, we want to quantify the efficiency gap between their corresponding social costs. Define the price of anarchy (PoA) as the maximum ratio between the social cost in (\ref{Cm}) under the myopic policy and the minimum cost in (\ref{C*}) under the socially optimal policy \cite{roughgarden2005selfish}:
\begin{align}
    \text{PoA}^{(m)}=\max_{\begin{aligned}&\Scale[0.65]{\alpha_H,\alpha_L,\mathbf{x}(t),\ell_0,N(t)}\\[-6pt]&\Scale[0.65]{\ \ \mathbf{L}(t),q_H,q_L,V(\cdot)}\end{aligned}} \frac{{C^{(m)}\big(\mathbf{L}(t),\mathbf{x}(t),N(t)\big)}}{{C^*\big(\mathbf{L}(t),\mathbf{x}(t),N(t)\big)}},\label{PoAm}
\end{align}
by searching all possible parameters and error function $V(\cdot)$ of the dynamic system. It is obvious that ${\text{PoA}^{(m)}\geq 1}$. 

In the next theorem, we prove the lower bound of PoA. 
\begin{theorem}\label{thm:PoAm}
In the parallel traffic network with $M$ stochastic paths, as compared to the minimum social cost in (\ref{C*}), the myopic policy in (\ref{Cm}) results in: 
\begin{align}
    \text{PoA}^{(m)}\geq \frac{{2(1-\rho^\Psi)}}{{2-\rho-\rho^\Psi}},\label{PoA>2}
\end{align}
where $\Psi=1+\log_{\alpha_H}\left(M\frac{\big(\ell_0-\frac{N_{min}}{M}-V(\frac{N_{min}}{M})\big)(\alpha_H-1)}{\alpha_H N_{max}}+1\right)$. The lower bound in (\ref{PoA>2}) is larger than $2$ as $\rho\rightarrow 1$, $V(\frac{N_{min}}{M})\ll\ell_0$ and $\ell_0\gg \alpha_H \frac{N_{max}}{M}$.
\end{theorem}

The proof of Theorem \ref{thm:PoAm} is given in Appendix E of the supplement material. Inspired by Proposition \ref{Prop:explore}, the worst case may happen when the myopic policy maximally under-explores stochastic paths with strong hazard belief $x_i(t)$. Initially, we set expected correlation coefficient ${\mathbb{E}[\alpha_i(0)|x_i(0)]=1}$ and expected travel costs ${c_i(0)=c_0(N_{max})}$ for any stochastic path $i$. Then according to (\ref{nm(t)}), myopic users always choose safe path 0 to make ${n_i^{(m)}(t)=0}$. However, the socially optimal policy frequently asks ${n_i^*(t)>0}$ of new user arrivals to explore stochastic path~$i$ to learn ${\alpha_L=0}$, which greatly reduces future travel latency on this path. After that, all users will keep exploiting path~$i$ with $\alpha_i(t)=\alpha_L$, and the expected travel cost on this path may gradually increase to ${c_i(\frac{N_{min}}{M})=c_0(0)}$ again after at least $\Psi$ time slots. As ${\Psi\rightarrow \infty}$ (by setting small variance cost ${V(\frac{N_{min}}{M})\ll\ell_0}$ and large latency ${\ell_0\gg\alpha_H \frac{N_{max}}{M}}$) and $\rho \rightarrow 1$, we have ${\text{PoA}^{(m)}\geq 2}$.

We can also show that the $\text{PoA}^{(m)}$ lower bound in (\ref{PoA>2}) decreases with variance cost ${V(\cdot)}$. This is consistent with Lemma \ref{lemma:n*-nm} and Proposition \ref{Prop:explore} that $n_i^{(m)}(t)$ approaches to ${n_i^*(t)}$ as ${V(\cdot)}$ enlarges. Besides, if stochastic path number ${M\rightarrow \infty}$, $\Psi$ approaches infinity, and the lower bound of PoA in (\ref{PoA>2}) becomes $2$, as more stochastic paths can negate the congestion cost caused by users' exploration.
As users' myopic routing can at least double the social cost with ${\text{PoA}^{(m)}\geq 2}$, we are well motivated to design an efficient mechanism to regulate.

\section{CHAR Mechanism with Learning Convergence}\label{section5}
In this section, we aim to design an informational mechanism to regulate users' myopic routing. We first analyze $x_i(t)$'s learning convergence under both myopic and socially optimal policies. Based on this analysis, we further show that existing information-hiding and deterministic-recommendation mechanisms are far from efficient. Accordingly, we propose our combined hiding and probabilistic recommendation (CHAR) mechanism and prove its close-to-optimal efficiency.

\subsection{Learning Convergence Analysis}

In the next lemma, we show that the myopic policy fails to ensure correct long-run convergence of hazard belief $x_i(t)$.

\begin{lemma}\label{lemma:myopic_learning}
Under the myopic policy with exploration number $n^{(m)}_i(t)$ on path $i$, the learned $x_i(t)$ is not guaranteed to converge to its real steady state $\Bar{x}$ in (\ref{bar_alpha}) as $t\rightarrow \infty$.
\end{lemma}

The proof of Lemma 4 is given in Appendix F of the supplement material. Under the myopic policy, users only care about the immediate travel latency of each path. If the travel latencies of stochastic paths all satisfy ${\mathbb{E}[\ell_i(t)|x_i(t)]> \ell_0}$ with ${\mathbb{E}[\alpha_i(t)|x_i(t)]\geq 1}$, myopic user arrivals will never explore any stochastic path $i$ and fail to update $x_i(t)$. In this case, hazard belief ${x_i(t)}$ may deviate a lot from its real steady state ${\Bar{x}}$. The deviation of $x_i(t)$ also incurs extra costs to mislead future users. Next, we examine the learning convergence of ${x_i(t)}$ under the socially optimal policy.

\begin{proposition}\label{Prop:optimal_stationary}
Under the socially optimal policy with exploration number $n_i^*(t)$, once ${V(N_{min})<\ell_0}$, as ${t\rightarrow \infty}$, ${x_i(t)}$ is guaranteed to converge to its real steady state ${\Bar{x}}$ in (\ref{bar_alpha}).
\end{proposition}

The proof of Proposition 2 is given in Appendix G of the supplement material. If ${V(N_{min})\geq \ell_0}$, the huge variance cost discourages the system to explore and converge to the real expected belief $\Bar{x}$ in any way. Note that even with ${V(N_{min})<\ell_0}$, the myopic policy is not guaranteed to converge to real expected belief $\Bar{x}$. While the socially optimal policy is willing to frequently explore stochastic path $i$ to learn a low-hazard state $\alpha_L$ to reduce the expected travel latency for the followers. With long-term frequent exploration, hazard belief $x_i(t)$ can finally converge to its real steady state $\Bar{x}$. The convergence to $\Bar{x}$ also helps the system best decide routing and reduce social costs.

\begin{figure}[t]
    \centering
    \includegraphics[width=0.38\textwidth]{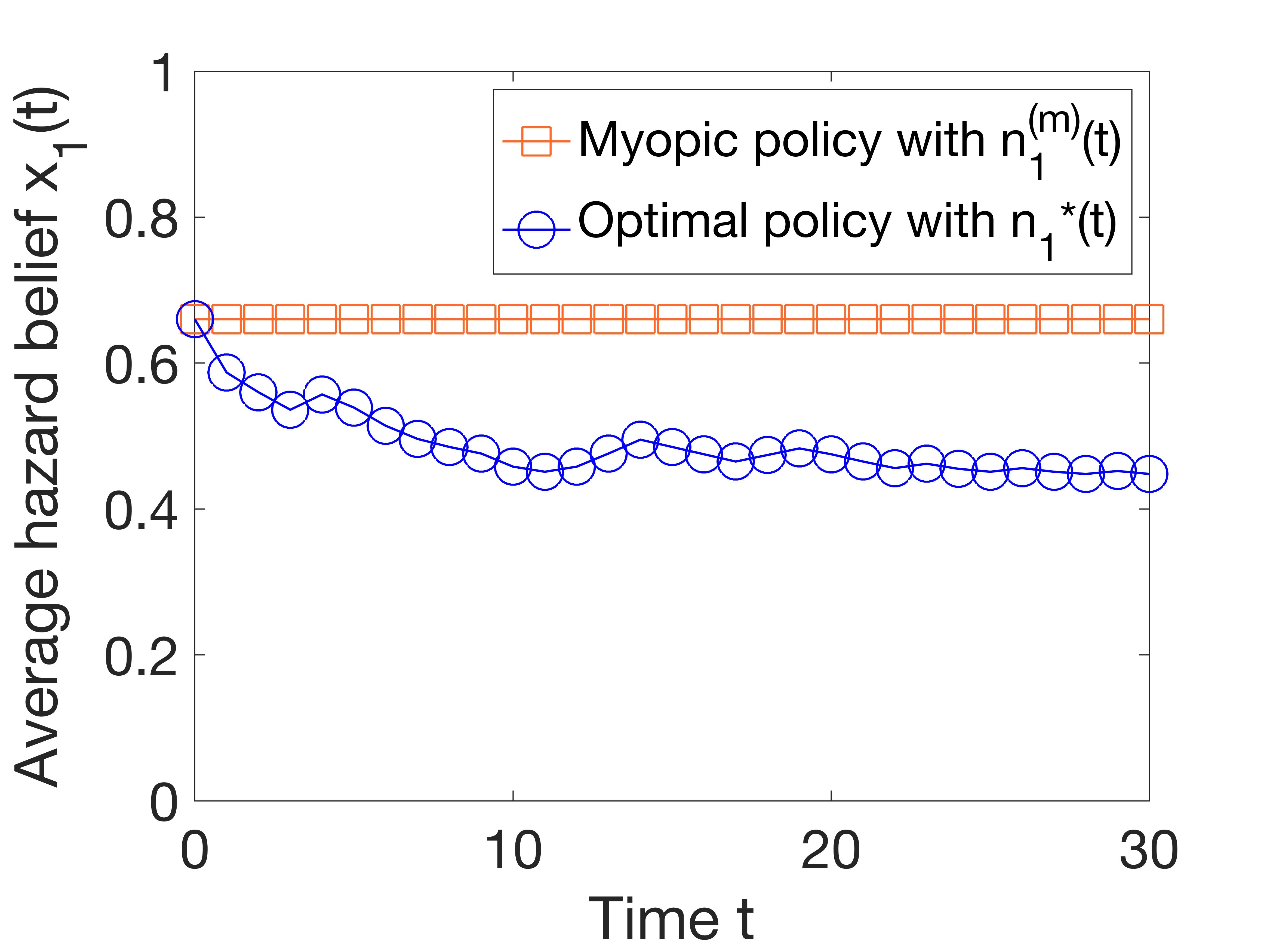}
    \captionsetup{font={footnotesize}}
    \caption{Dynamics of average hazard belief $x_1(t)$ under both myopic and socially optimal policies from $t=0$ to $30$ in a two-path network with $M=1$.}
    \label{fig:xbar}
\end{figure}

In Fig. \ref{fig:xbar}, we run multiple experiment instances to show the average convergence of ${x_1(t)}$ under both policies in a finite time horizon ${\{0,\cdots,30\}}$. Here we set $N(0)=\Bar{N}=7,\ell_0=100,x_1(0)=0.66,\mathbb{E}[\ell_1(0)|x_1(0)=0.6]=110,n_1(0)=0,\rho=0.99,\alpha_H=1.5,\alpha_L=0.05$, and $\Bar{\alpha}=0.7$ with $\Bar{x}=0.45$ at initial time $t=0$. Here $q_H(n_1(t))$ in (\ref{F1F0}) follows a normal distribution's CDF with mean $0.3$ and variance $1$, and we properly choose error function $V(n(t-1))=\min\{\frac{10}{n(t-1)},20\}$. Under these parameter settings, the myopic policy never explores stochastic path 2, because the expected latency on path 2 keeps increasing under ${\mathbb{E}[\alpha(t)|x(t-1)]=1.007}$. Hence, the myopic policy makes $x(t)=0.66$ deviate from real but unknown ${\Bar{x}=0.45}$ in Fig. \ref{fig:xbar}. However, the socially optimal policy still recommends some users to learn useful information on path 2, which ensures $x(t)$'s correct convergence to $\Bar{x}=0.45$. The simulation results above are consistent with Lemma \ref{lemma:myopic_learning} and Proposition \ref{Prop:optimal_stationary}.

\subsection{Benchmark Informational Mechanisms Comparison}\label{section:benchmark}

Before presenting our new mechanism, we first follow the mechanism design literature \cite{fudenberg1991game} to generally define the informational mechanism below. 
\begin{definition}[Informational mechanisms]\label{def:informational_mechanism}
An informational mechanism defines an incomplete information Bayesian game, where the social planner determines the disclosing level of variables based on the system state. 
\end{definition}

In practice, informational mechanisms are non-monetary and easier to implement as compared to pricing and side-payment mechanisms (e.g., \cite{zheng2020designing,Ferguson2022effective,li2024incentivizing}).
Then we analyze the efficiency of two widely used informational mechanisms: information-hiding \cite{tavafoghi2017informational,wang2020efficient,farhadi2022dynamic} and deterministic-recommendation \cite{li2019recommending,wu2019learning,li2023congestion}. Note that we also tried prior Bayesian persuasion mechanisms \cite{das2017reducing,kamenica2019bayesian,mansour2022bayesian}, which focus on one-shot games and cannot optimize long-run information learning.

Based on Proposition \ref{Prop:optimal_stationary}, if the platform hides all the information, i.e., $\mathbf{L}(t),\mathbf{x}(t)$ and $N(t)$, as \cite{tavafoghi2017informational,wang2020efficient,farhadi2022dynamic}, users can only estimate that each stochastic path has reached its steady state ${\Bar{x}\sim \mathbb{P}(\Bar{x})}$ in (\ref{bar_alpha}) and ${\mathbb{E}[N(t)]=\Bar{N}}$ for any time $t$. As in \cite{tavafoghi2017informational}, expected exploration number $n_i^{\emptyset}\in\{0,1,\cdots,N(t)\}$ of stochastic path $i$ under information-hiding becomes constant:
\begin{align}
    n_i^{\emptyset}=\min
    \left\{\frac{N(t)}{{M}},\frac{\Bar{N}+c_0(0)-\mathbb{E}_{\Bar{x}\sim \mathbb{P}(\Bar{x})}[c_i(0)|\Bar{x}]}{M+1}\right\}, \label{n_empty}
\end{align}
where ${c_0(\cdot)}$ and ${c_i(\cdot)}$ are defined in (\ref{cm_1}) and (\ref{cm_2}), respectively.
We next prove the infinite PoA caused by the hiding mechanism.
\begin{lemma}\label{lemma:information_hiding}
If the platform hides all the information, i.e., $\mathbf{L}(t),\mathbf{x}(t)$ and $N(t)$, from users, the constant policy $n^{\emptyset}_i$ in (\ref{n_empty}) makes users either under- or over-explore stochastic path~$i$ as compared to the social optimum, leading to $\text{PoA}=\infty$.
\end{lemma}

The proof of Lemma 5 is given in Appendix H of the supplement material. If ${\mathbb{E}_{\Bar{x}\sim \mathbb{P}(\Bar{x})}[c_i(\frac{N_{max}}{M})|\Bar{x}]\leq c_0(0)}$, all the ${N(t)}$ users without information will choose stochastic paths with less expected cost. However, the actual ${\alpha_i(t)}$ can be ${\alpha_H}$ and ${\mathbb{E}[\ell_i(t)|x_i(t)]\gg \ell_0}$, then users’ maximum over-exploration makes PoA arbitrarily large. Similarly, if ${\mathbb{E}_{\Bar{x}\sim \mathbb{P}(\Bar{x})}[c_i(0)|\Bar{x}]> c_0(N_{max})}$, all users will choose path 0, leading to the maximum under-exploration.

Next, we consider the existing deterministic recommendation mechanisms (\!\!\cite{li2019recommending,wu2019learning,li2023congestion}), which privately provide state-dependent deterministic recommendations to users while hiding other information, to prove the caused infinite PoA. 
\begin{lemma}\label{lemma:deterministic_recommend}
The deterministic-recommendation mechanism makes $\text{PoA}=\infty$ in our system.
\end{lemma}

The proof of Lemma 6 is given in Appendix I of the supplement material. When a user receives recommendation ${\pi(t)=i}$, he only infers ${n_i(t)\geq 1}$ on path $i$. However, if expected user number $\Bar{N}\gg 1$, this information is insufficient to alter his posterior distribution of ${\Bar{x}}$ from ${\mathbb{P}(\Bar{x})}$. Consequently, the caused exploration number still approaches ${n^{\emptyset}_i}$ in (\ref{n_empty}), leading to ${\text{PoA}=\infty}$. 

\subsection{New CHAR Mechanism Design and Analysis}

Inspired by Subsection \ref{section:benchmark}, the previous two mechanisms do not work in improving our system performance. Thus, we will properly combine them for a new mechanism design. To approach optimal policy $n_i^*(t)$ as much as possible, we dynamically select a number ${N^{\emptyset}(t)}$ of users to follow hiding policy $n_i^{\emptyset}$ in (\ref{n_empty}), while providing state-dependent probabilistic recommendations to the remaining ${N(t)-N^{\emptyset}(t)}$ users.

Now we are ready to propose our combined hiding and probabilistic recommendation (CHAR) mechanism, which contains two steps per time slot $t$. Under CHAR, let ${n^{(\text{CHAR})}_i(t)\in\{0,\cdots,N(t)\}}$ represent the exploration number of path $i$ at $t$. Similar to (\ref{Cm}), denote ${C^{(\text{CHAR})}(\mathbf{L}(t),\mathbf{x}(t),N(t))}$ to be the long-term social cost under $\mathbf{n}^{(\text{CHAR})}(t)$.
\begin{definition}[CHAR mechanism] \label{def:CHAR}
At any time $t$, in the first step, the platform randomly divides the ${N(t)}$ users into the hiding-group with ${N^{\emptyset}(t)}$ users and the recommendation-group with ${N(t)-N^{\emptyset}(t)}$ users, where the dynamic number ${N^{\emptyset}(t)}$ is the optimal solution to 
\begin{align}
    \min_{N^{\emptyset}(t)\in\{0,\cdots,N(t)\}} &{C^{(\text{CHAR})}(\mathbf{L}(t),\mathbf{x}(t),N(t))},\label{Cc}\\
    \text{s.t.} \quad\quad\ \ &{n_i^{(\text{CHAR})}(t)=N^{\emptyset}(t)}\cdot\frac{{n_i^{\emptyset}}}{{N(t)}}+\label{n^*:N_H}\\ &\quad\quad\quad\ \ {(N(t)-N^{\emptyset}(t))\mathbf{Pr}\big(\pi(t)=i|x_i(t)\big)}\notag,
\end{align}
where
\begin{align}
    {\mathbf{Pr}\big(\pi(t)=i|x_i(t)\big)=}\begin{cases}
        {p_L},\ \text{ if }{x_i(t)< x_{th}},\\
        {p_H},\text{ if }{x_i(t)\geq x_{th}},
    \end{cases}\label{Pr(pi=2)}
\end{align}
with $x_{th}$ is in Proposition \ref{Prop:explore} and any feasible ${p_L}$ and ${p_H}$ to satisfy $p_L\mathbb{P}(x_{th})\geq p_H(1-\mathbb{P}(x_{th}))$.

In the second step, the platform performs as follows:
\begin{itemize}
    \item For the ${N^{\emptyset}(t)}$ hiding-group users, the platform hides all the past information from them. 
    \item For the rest ${N(t)-N^{\emptyset}(t)}$ users in the recommendation-only group, the platform randomly recommends a path $\pi(t)$ to each user arrival by the following distribution:
    \begin{align}
        \pi(t)=\begin{cases}
            i,&\text{with }{\mathbf{Pr}(\pi(t)=i|x_i(t))} \text{ in (\ref{Pr(pi=2)})},\\
            0,&\text{with }{1-\sum_{i=1}^M\mathbf{Pr}(\pi(t)=i|x_i(t))}.
        \end{cases}\label{recommend_pi}
    \end{align}
\end{itemize}
\end{definition}

According to Definition \ref{def:CHAR}, the $N^{\emptyset}(t)$ users in the uniformly selected hiding-group rely on hiding policy $n_i^{\emptyset}$ in (\ref{n_empty}) to make routing decisions. For the recommendation-only group, given the always feasible $p_L$ and $p_H$ to satisfy $p_L \mathbb{P}(x_{th})\geq p_H (1-\mathbb{P}(x_{th}))$, if a user receives a recommendation $\pi(t)=i$, he infers that the posterior belief of low hazard $x_i(t)<x_{th}$ on path $i$ is larger than the high harzard $x_i(t)\geq x_{th}$. Thus, this user will not deviate from the recommendation for a smaller expected travel cost. Similarly, if recommended $\pi(t)=0$, the user can infer that any stochastic path $i$ is more likely to have $x_i(t)\geq x_{th}$ and accordingly will choose safe path 0. 

Given Bayesian incentive compatibility for all users, according to (\ref{n^*:N_H}), there are respectively ${N^{\emptyset}(t)\cdot n_i^{\emptyset}/N(t)}$ users in the hiding-group and ${(N(t)-N^{\emptyset}(t))\cdot \mathbf{Pr}(\pi(t)=i|x_i(t))}$ users in the recommendation-only group traveling on any path~$i$. 
If stochastic path $i$ has a good condition with $x_i(t)<x_{th}$ at time $t$, the recommendation probability in (\ref{Pr(pi=2)}) is $\mathbf{Pr}\big(\pi(t)=i|x_i(t)\big)=p_L$. Then the exploration number of users on path~$i$ under our CHAR mechanism satisfies
\begin{align*}
    n_i^{(\text{CHAR})}=N^{\emptyset}(t)\Big(\frac{n_i^{\emptyset}}{N(t)}-p_L\Big)+p_LN(t),
\end{align*}
which depends on the choice of $N^{\emptyset}(t)\in\{0,\cdots, N(t)\}$. While if path~$i$ has a bad condition with $x_i(t)\geq x_{th}$, $n_i^{(\text{CHAR})}$ is similarly derived.
To make $n_i^{(\text{CHAR})}(t)$ approach optimum $n_i^*(t)$ on each path~$i$, the platform optimizes (\ref{Cc}) to determine the optimal solution $N^{\emptyset}(t)$ in each time slot~$t$.

Then we propose the following theorem to prove the close-to-optimal performance of our CHAR mechanism.
\begin{theorem}\label{thm:poa=1}
Our CHAR mechanism in Definition \ref{def:CHAR} ensures Bayesian incentive compatibility for all users and regulates the dynamic routing system to achieve 
\begin{align}
    {\text{PoA}^{(\text{CHAR})}=1+\frac{1}{2(M+1)\big(1+\frac{M}{\Bar{N}}\cdot V\big(\frac{\Bar{N}(2M+1)}{2M(M+1)}\big)\big)}},\label{PoA_c}
\end{align}
which is always less than $\frac{5}{4}$ and cannot be further reduced by any informational mechanism.
\end{theorem}

\begin{figure*}[t]
    \centering
    \includegraphics[width=0.85\textwidth]{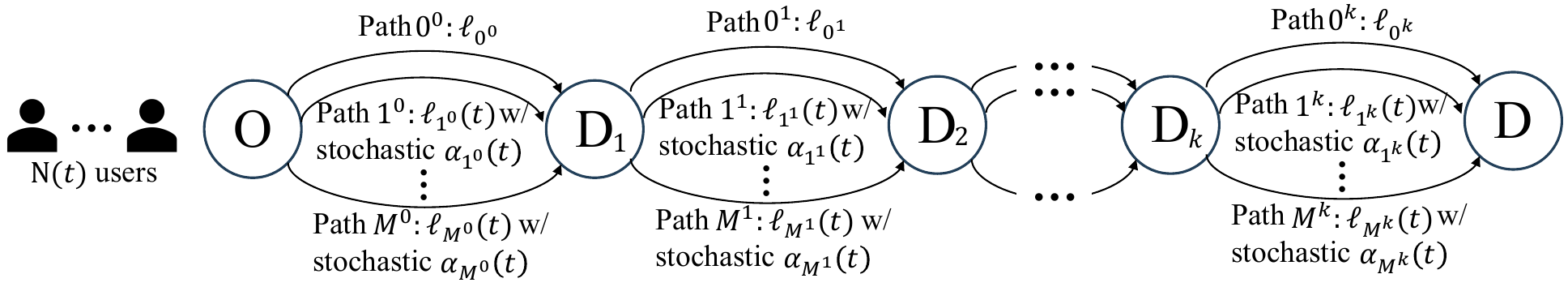}
    \captionsetup{font={footnotesize}} 
    \caption{Generalization from the parallel graph in Fig.~\ref{fig:congestion_game} to a linear path graph: At the beginning of each time slot $t\in\{1,2,\cdots\}$, $N_j(t)$ users arrive at any node $\text{D}_j\in\{\text{O},\text{D}_1,\cdots, \text{D}_k\}$ selects one path from $M+1$ available paths to travel to the next node in this linear path graph. Among all the $M+1$ paths between intermediate nodes $\text{D}_{j}$ and $\text{D}_{j+1}$, where $j\in\{0,\cdots,k\}$, path $0^j$ is safe while any path $i^j\in\{1^j,\cdots,M^j\}$ is stochastic.}
    \label{fig:congestion_game_k}
\end{figure*}

The proof of Theorem \ref{thm:poa=1} is given in Appendix J of the supplement material. On one hand, our CHAR mechanism recommends $n_i^{(\text{CHAR})}>n_i^{(m)}(t)$ users to explore stochastic path~$i$ of high belief $x_i(t)\geq x_{th}$. This successfully avoids myopic policy's zero-exploration in Theorem \ref{thm:PoAm}. On the other hand, our CHAR well controls the number of users on stochastic paths with bad conditions, avoiding benchmark mechanisms' maximum exploration in bad traffic conditions in Lemmas~\ref{lemma:information_hiding} and \ref{lemma:deterministic_recommend}. Therefore, it efficiently reduces ${\text{PoA}^{(m)}\geq 2}$ caused by the myopic policy. Under our CHAR, the worst case occurs when users over-explore stochastic paths in good conditions ($\alpha_i(t)=\alpha_L$).
We consider the scenario where the travel cost on any stochastic path $i$ satisfies $c_i\big(\frac{N_{max}}{M}\big)<c_0(0)$ even if $\alpha_i(t)=\alpha_H$. Then all users will opt for stochastic paths, and no information incentives can curb their over-exploration. The achieved minimum PoA is derived in (\ref{PoA_c}), which cannot be further reduced by any informational mechanism.

Our CHAR's PoA in (\ref{PoA_c}) increases with the expected user number $\Bar{N}$ and decreases with path number ${M}$. As ${\Bar{N}\rightarrow \infty}$, ${\text{PoA}^{(\text{CHAR})}}$ converges to $1+\frac{1}{2(M+1)}$ with minimum $\frac{5}{4}$. While if ${M\rightarrow\infty}$, ${\text{PoA}^{(\text{CHAR})}}$ approaches the optimum $1$.

\section{Extensions of Parallel Transportation Network to Linear Path Graphs}\label{section6}
In this section, we extend our congestion model and analysis from the parallel transportation network between two nodes O and D in Fig.~\ref{fig:congestion_game} to more general linear path networks with multiple intermediate nodes and stochastic paths (\!\!\cite{Broersma1989pathgraph}). In this generalized model, we first establish that the PoA lower bound caused by the myopic policy remains unchanged. Next, we prove that our CHAR mechanism continues to effectively reduce the PoA to its minimum possible value.

As depicted in Fig. \ref{fig:congestion_game_k}, we consider a general linear path graph, which contains an order list of $k$ intermediate nodes, denoted by $\text{D}_1,\cdots, \text{D}_k$, between origin O and destination D. For ease of exposition, we denote $\text{D}_0=\text{O}$ and $\text{D}_{k+1}=\text{D}$. At each node $\text{D}_{j}$ for $j\in\{0,1,\cdots,k\}$, there exist one safe path~$0^{j}$ and $M$ stochastic paths $1^{j},\cdots, M^{j}$ leading to the next node $\text{D}_{j+1}$. The safe~path~$0^j$ within the $j$-th subnetwork has a fixed travel latency $\ell_{0^j}$. While for the other $M^j$ stochastic paths, their correlation coefficients still alternate between a high-hazard state $\alpha_H$ and a low-hazard state $\alpha_L$ over time.  

Within each subnetwork $j$ from node $\text{D}_j$ to $\text{D}_{j+1}$ of this generalized path network, users traveling on stochastic path $i^j$ will learn the traffic information there and update its hazard belief $x_{i^j}(t+1)$ for the platform as follows:
\begin{align*}
    x_{i^j}(t+1)=x'_{i^j}(t)q_{HH}(t)+(1-x'_{i^j}(t))q_{LH}(t).
\end{align*}

In this extended system model, a user arriving at node $\text{D}_j$ not only needs to consider his current travel cost $c_{i^j}(n_{i^j}(t))$ in the $j$-th subnetwork, but also cares about his cost-to-go in the future subnetworks $\{j+1,\cdots,k\}$. This makes the myopic policy different from the one-shot decision in (\ref{nm(t)}) for the parallel network in Fig. \ref{fig:congestion_game}. Nonetheless, we still managed to obtain the PoA caused by the myopic routing policy in the following proposition.

\begin{proposition}\label{prop:poa_k}
Under the general linear path graph in Fig.~\ref{fig:congestion_game_k}, as compared to the social optimum, the myopic policy achieves 
\begin{align*}
    \text{PoA}^{(m)}\geq \frac{{2(1-\rho^\Psi)}}{{2-\rho-\rho^\Psi}},
\end{align*}
where $\Psi=1+\log_{\alpha_H}\left(M\frac{\big(\ell_{0^j}-\frac{N_{min}}{M}-V(\frac{N_{min}}{M})\big)(\alpha_H-1)}{\alpha_H N_{max}}+1\right)$. The lower bound in (\ref{PoA>2}) is larger than $2$ as $\rho\rightarrow 1$, $V(\frac{N_{min}}{M})\ll\ell_{0^j}$ and $\ell_{0^j}\gg \alpha_H \frac{N_{max}}{M}$ for any $j\in\{0,\cdots,k\}$.
\end{proposition}

The proof of Proposition \ref{prop:poa_k} is given in Appendix K of the supplement material. In the linear path network illustrated in Fig.~\ref{fig:congestion_game_k}, we set an expected correlation coefficient $\mathbb{E}[\alpha_{i^j}(0)|x_{i^j}(0)]=1$ and expected travel costs ${c_{i^j}(0)=c_{0^j}(N_{max})}$ for any stochastic path $i^j$. In the worst-case scenario with $\rho=1$, the myopic policy still leads all users to choose the safe path $0^j$ at any node $\text{D}_j$. While the socially optimal policy always recommends $n_{i^j}^*(t)>0$ users to explore stochastic path $i^j$ to potentially learn possible $\alpha_L=0$ there. After that, all users keep exploiting stochastic path $i^j$ with $\alpha_{i^j}(t)=\alpha_L$, resulting in $\text{PoA}>2$. It is noteworthy that in this case, both the myopic policy and the socially optimal policy just repeat their decision-making processes, as in (\ref{nm(t)}) and (\ref{C*}), for $k+1$ times from origin O to destination D. Consequently, the resulting PoA remains the same as that of the parallel path network in Theorem~\ref{thm:PoAm}. 

Motivated by Proposition~\ref{prop:poa_k}, we next examine our CHAR mechanism's performance for this generalized system. In the following, we consider $\rho=1$ in the worst-case scenario to regulate myopic routing. Let vectors $\mathbf{L}^j(t)$ and $\mathbf{x}^j(t)$ summarize the latencies and beliefs of all the $M$ stochastic paths within the $j$-th subnetwork. Denote by $N^j(t)$ the arriving user number at node $\text{D}^j$ at time $t$. We first propose the following lemma to check users' decision-making without information sharing.
\begin{lemma}\label{lemma:hiding_k}
If the platform hides all the information of the linear path graph in Fig.~\ref{fig:congestion_game_k} from all users at any node $\text{D}_j$, i.e., $\mathbf{L}^j(t), \mathbf{x}^j(t)$ and $N^j(t)$ at any time $t$, the expected exploration number of stochastic path $i^j$ within the $j$-th subnetwork equals
\begin{align}\label{n_empty_k}
    n_{i^j}^{\emptyset}=\min
    \left\{\frac{N^j(t)}{M},\frac{\Bar{N}+c_{0^j}(0)-\mathbb{E}_{\Bar{x}\sim \mathbb{P}(\Bar{x})}[c_{i^j}(0)|\Bar{x}]}{M+1}\right\}.
\end{align}
\end{lemma}

The proof of Lemma \ref{lemma:hiding_k} is given in Appendix L of the supplement material. Intuitively, without any information, users can only estimate that each stochastic path $i^j$ has reached its steady state $\Bar{x}$ in (\ref{bar_alpha}) and $\mathbb{E}[N^j(t)]=\Bar{N}$ for any time~$t$. Under the steady state, for any user at node $\text{D}_j$, his expected cost-to-go for the future subnetworks $\{j+1,\cdots,k\}$ is unaffected by his current path choice. Therefore, he only needs to focus on the current subnetwork $j$, as (\ref{n_empty}) in the parallel network, to make his routing decision. This leads to the total exploration number $n_{i^j}^{\emptyset}$ in (\ref{n_empty_k}) for path $i^j$.

Thanks to Lemma~\ref{lemma:hiding_k}, we only need to focus on a single subnetwork $\{\text{D}_j,\text{D}_{j+1}\}$ to regulate each user arrival at node $\text{D}_j$. As a result, our CHAR mechanism for the parallel network in Definition \ref{def:CHAR} still works. Subsequently, we prove its PoA result in the following theorem.

\begin{theorem}\label{thm:poa=1_k}
Under the linear path graph in Fig.~\ref{fig:congestion_game_k}, our CHAR mechanism in Definition \ref{def:CHAR} regulates the dynamic system to achieve the same PoA as (\ref{PoA_c}), which is always less than $\frac{5}{4}$.
\end{theorem}

The proof of Theorem \ref{thm:poa=1_k} is given in Appendix M of the supplement material. Since our CHAR mechanism only needs to regulate users within each subnetwork $j$, the achieved PoA within each subnetwork is equal to (\ref{PoA_c}), as in the parallel transportation network. When considering all $k+1$ subnetworks, the PoA remains unchanged.

\section{Experiment Validation Using Real Datasets}\label{section7}
\begin{figure}[t]
    \centering
    \includegraphics[width=0.48\textwidth]{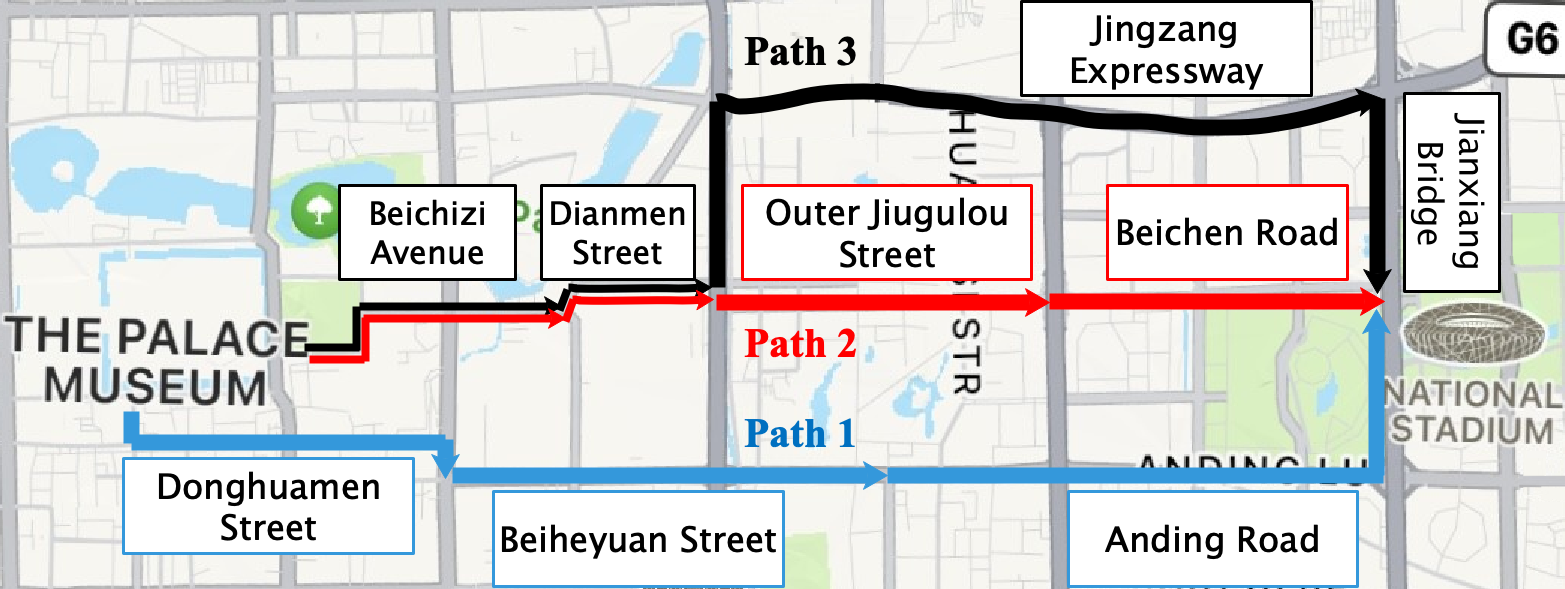}
    \captionsetup{font={footnotesize}}
    \caption{A popular hybrid road network consisting of three path choices from the Palace Museum to the National Stadium in Beijing, China. In this hybrid network, each path has a series of roads, where Path 2 and Path 3 overlap on Beichizzi Avenue and Dianmen Street, making it different from the linear path graph in Fig. \ref{fig:congestion_game_k}. On public holidays, many visitors randomly leave the Palace Museum and choose a path to travel to the National Stadium.}
    \label{fig:map}
\end{figure}

In addition to the worst-case PoA analysis, in this section, we further conduct experiments to evaluate our CHAR's average performance versus the myopic policy used by Waze and Google Maps and the popular information hiding mechanism used in existing congestion game literature (\!\!\cite{tavafoghi2017informational,farhadi2022dynamic,wang2020efficient}). To further practicalize our dynamic congestion model in (\ref{ell_2}), we sample peak hours' real-time traffic congestion data in Beijing, China on public holidays using BaiduMap (\!\!\cite{Baidumap}), and further extend our parallel transportation network in Fig.~\ref{fig:congestion_game} and linear path graph in Fig.~\ref{fig:congestion_game_k} to a hybrid road network in Fig.~\ref{fig:map}. 

As depicted in Fig. \ref{fig:map}, we consider a popular hybrid road network from the Palace Museum to the National Stadium on public holidays, as many visitors randomly leave the Palace Museum and travel to the National Stadium. This hybrid network consists of three path choices, each with a series of safe and stochastic roads:
\begin{itemize}
    \item Path 1: Donghuamen Street, Beiheyuan Street, and Anding Road.
    \item Path 2: Beichizi Avenue, Dianmen Street, Outer Jiugulou Street and Beichen Road.
    \item Path 3: Beichizi Avenue, Dianmen Street, Jingzang Expressway, and Jianxiang Bridge.
\end{itemize}
Here Path 2 and Path 3 overlap on Beichizzi Avenue and Dianmen Street, making it different from the linear path graph in Fig. \ref{fig:congestion_game_k}.

To train a practical congestion model with travel latency, we mine and analyze many data about the real-time traffic status values of the nine roads from \cite{Baidumap}, which can dynamically change every $5$ minutes in the peak-hour periods. We validate from the dataset that the traffic conditions of Donghuamen Street and Beiheyuan Street on Path 1, Beichizi Avenue on both Path 2 and Path 3, and Jianxiang Bridge on Path 3 can be well approximated as Markov chains with two discretized states (high and low traffic states) as in (2), while the other five roads tend to have deterministic/safe conditions. Similar to \cite{eddy1998profile,chen2016predicting}, we employ the hidden Markov model (HMM) approach to train the transition probability matrices for the four stochastic road segments using $432$ statuses with $5$ minutes per time slot, which suffice to achieve high accuracy. Based on the transition probability matrices, we further calculate the steady-state belief $\Bar{x}$'s of the four stochastic roads in Fig. \ref{fig:map}:
\begin{align*}
    \Bar{x}_{\text{Donghuamen}}&=0.388,\ 
    \Bar{x}_{\text{Beihe}}\ \ \ =0.106,\\
    \Bar{x}_{\text{Beichizi}}&=0.192,\ 
    \Bar{x}_{\text{Jianxiang}}=0.936.
\end{align*}
Then these steady-state beliefs can be used in the congestion model for our experiments.

Besides the above congestion model, based on the travel latency data in \cite{Baidumap}, we calculate the long-term average correlation coefficients $\alpha_H\approx 1.3$ and $\alpha_L\approx 0.3$, by assuming a linear latency function $f(\cdot)$ in (\ref{ell_2}). Then we use the traffic flow data from Baidu Map to calculate the mean arrival car number is $N=121$ every $5$ minutes at the gateway of the Palace Museum, with a standard deviation of $12.33$. Given the mean arrival car number, we set $V(n(t-1))=\min\{\frac{100}{n(t-1)},200\}$.

Based on the derived system parameters above, we are now ready to conduct our experiments. In the myopic policy, each user estimates the travel cost of any path choice, by aggregating the costs of its road segments, to select the one that minimizes his own travel cost, whereas the socially optimal policy seeks the path choice that minimizes the collective long-term social cost for all users. Our CHAR mechanism optimizes the user number of the hiding-group $N^{\emptyset}(t)$ in (\ref{Cc}) and recommends path $\pi(t)$ to each user of the recommendation-only group in (\ref{recommend_pi}). This is achieved by combining the actual hazard beliefs of all stochastic road segments on each path choice. As users' costs are proportional to their travel delays, we conduct $50$ experiments, each with $31$ time slots (equivalent to $155$ minutes), to calculate their average long-term social costs (in minutes). Here the initial hazard beliefs are approximated to $0.5,0.2,0.3$, and $0.8$ for Donghuamen Street, Beiheyuan Street, Beichizi Avenue, and Jianxiang Bridge, respectively.

\begin{figure}[t]
    \centering
    \includegraphics[width=0.38\textwidth]{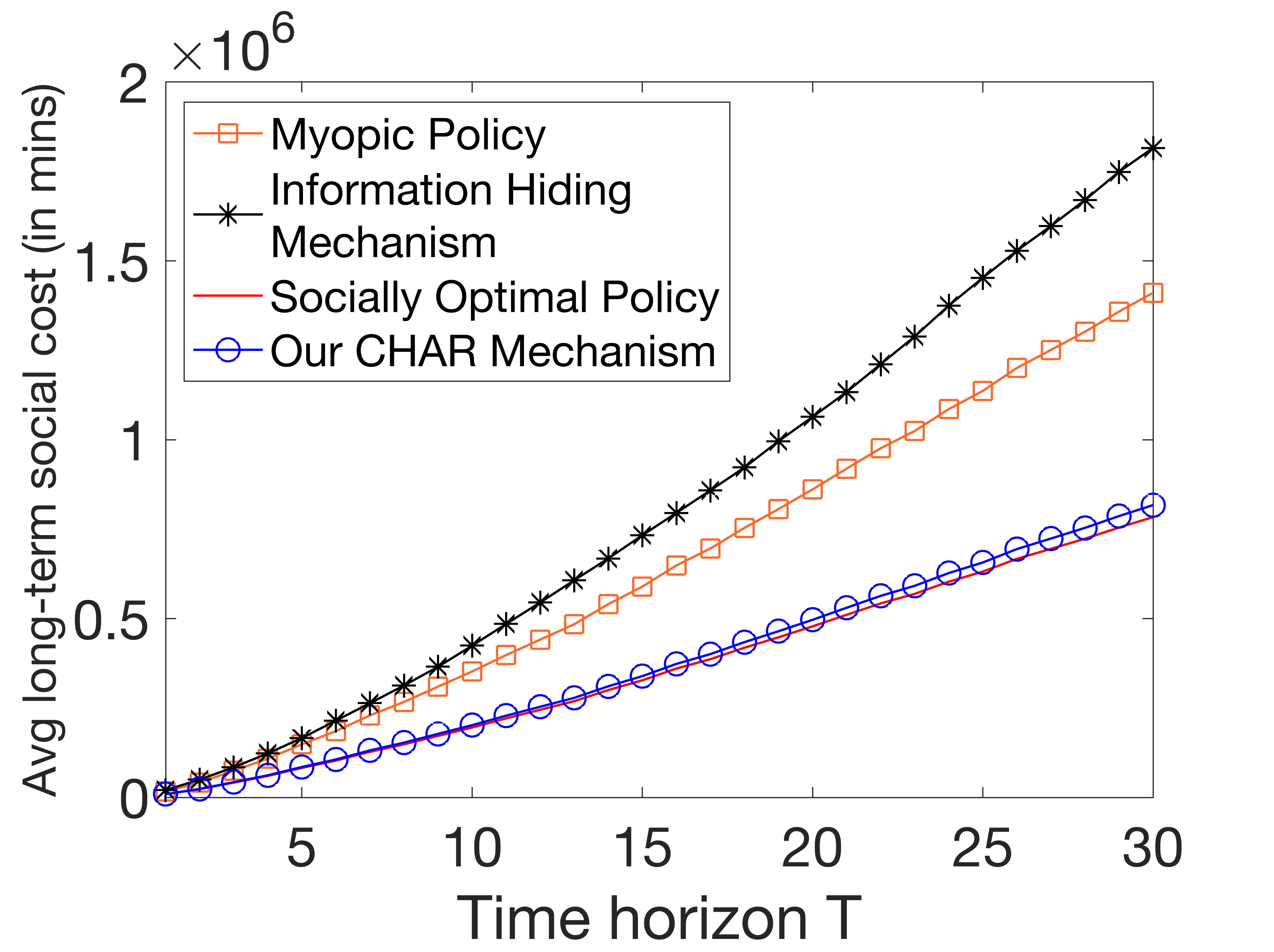}
    \captionsetup{font={footnotesize}}
    \caption{Average long-term costs (in minutes) under the myopic policy, information hiding mechanism (\!\!\cite{tavafoghi2017informational,farhadi2022dynamic,wang2020efficient}), the socially optimal policy, and our CHAR mechanism versus time horizon $T\in\{0,\cdots,30\}$.}
    \label{fig:avg_cost}
\end{figure}

In the first experiment, we fix discount factor $\rho=0.98$ to compare the average long-term social costs under the myopic policy, information hiding mechanism, the socially optimal policy, and our CHAR mechanism versus time horizon $T\in\{0,\cdots, 30\}$. Fig. \ref{fig:avg_cost} depicts that our CHAR mechanism has less than $5\%$ efficiency loss from the social optimum for any time horizon $T$, while the myopic policy and the information hiding mechanism cause around $90\%$ and $125\%$ efficiency losses, respectively.

\begin{figure}[t]
    \centering
    \includegraphics[width=0.38\textwidth]{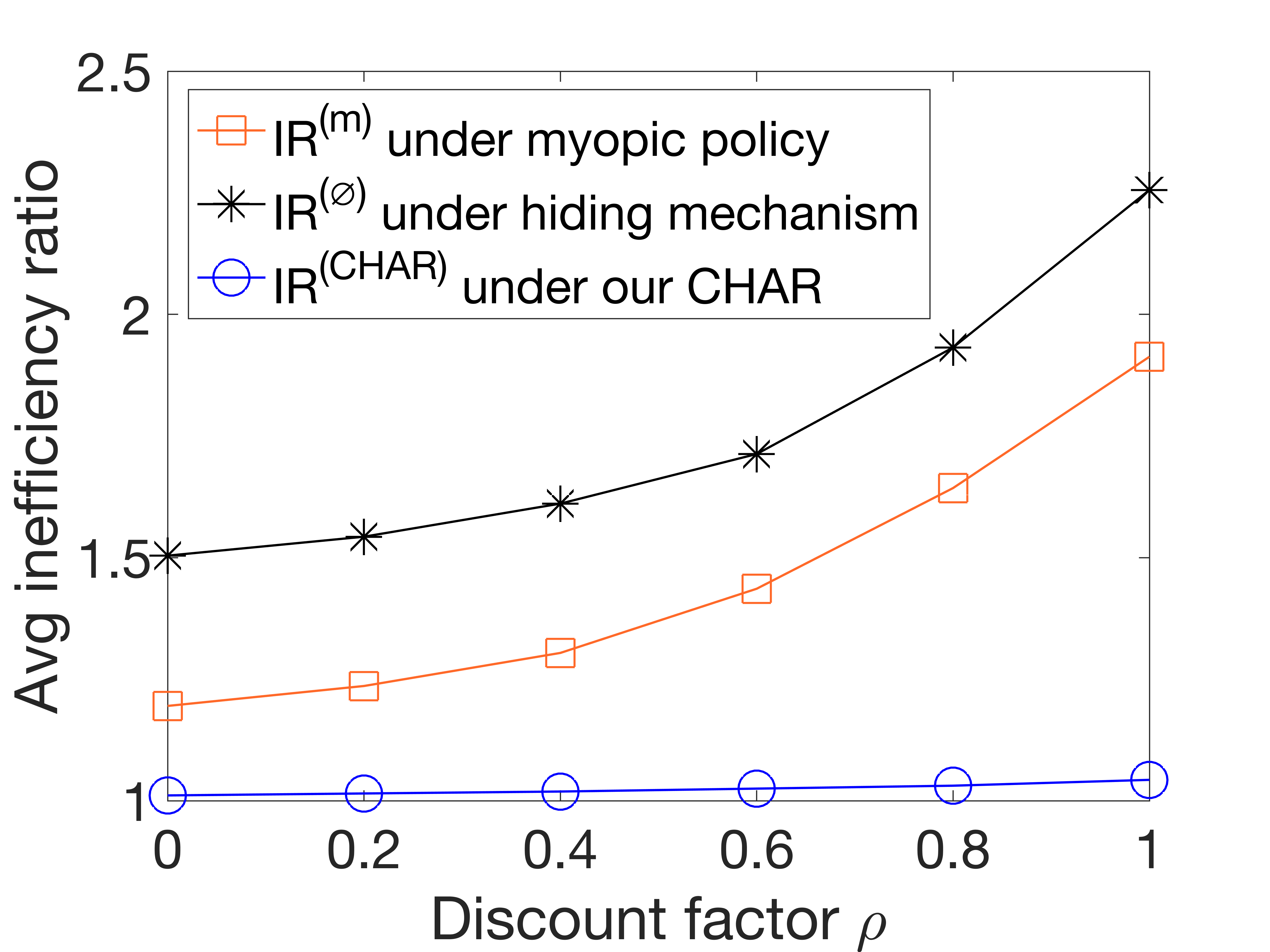}
    \captionsetup{font={footnotesize}}
    \caption{Average long-term inefficiency ratios $\text{IR}^{(m)}$, $\text{IR}^{(\emptyset)}$, and $\text{IR}^{(\text{CHAR})}$ caused by the myopic policy, information-hiding mechanism, and our CHAR mechanism (as compared to the socially optimal policy) versus discount factor~$\rho\in\{0,0.2,0.4,0.6,0.8,1\}$.}
    \label{fig:avg_ineff}
\end{figure}

Next, we define 
\begin{align*}
    \text{IR}^{(m)}=\frac{\mathbb{E}[C^{(m)}(\mathbf{L}(t),\mathbf{x}(t),N(t))]}{\mathbb{E}[C^*(\mathbf{L}(t),\mathbf{x}(t),N(t))]}
\end{align*}
to be the long-term average inefficiency ratio of the total travel cost $C^{(m)}(\mathbf{L}(t),\mathbf{x}(t),N(t))$ under the myopic policy in (\ref{Cm}), as compared to $C^*(\mathbf{L}(t),\mathbf{x}(t),N(t))$ under the socially optimal policy in (\ref{C*}). Similarly, define 
\begin{align*}
    \text{IR}^{(\emptyset)}&=\frac{\mathbb{E}[C^{(\emptyset)}(\mathbf{L}(t),\mathbf{x}(t),N(t))]}{\mathbb{E}[C^*(\mathbf{L}(t),\mathbf{x}(t),N(t))]},\\
    \text{IR}^{(\text{CHAR})}&=\frac{\mathbb{E}[C^{(\text{CHAR})}(\mathbf{L}(t),\mathbf{x}(t),N(t))]}{\mathbb{E}[C^*(\mathbf{L}(t),\mathbf{x}(t),N(t))]}
\end{align*}
to be the average inefficiency ratios caused by the information hiding mechanism and our CHAR mechanism, respectively. Then in the second experiment, we fix time horizon $T=30$ to compare the average inefficiency ratios $\text{IR}^{(m)}, \text{IR}^{(\emptyset)}$ and $\text{IR}^{(\text{CHAR})}$ caused by the myopic policy, information hiding mechanism, and our CHAR mechanism versus discount factor $\rho\in\{0,0.2,0.4,0.6,0.8,1\}$. 

Fig.~\ref{fig:avg_ineff} tells that all the inefficiency ratios increase with discount factor $\rho$, which is aligned with Theorem~\ref{thm:PoAm}. In the worst case with $\rho=1$, our CHAR mechanism obviously reduces $\text{IR}^{(m)}>1.9$ and $\text{IR}^{\emptyset}>2.2$ under the myopic policy and hiding mechanism to the close-to-optimal ratio $\text{IR}^{(\text{CHAR})}<1.1$, which is consistent with Theorem \ref{thm:poa=1}. Note that even in the best case with $\rho=0$, both myopic policy and information hiding mechanism perform poor with $\text{IR}^{(m)}>1.2$ and $\text{IR}^{\emptyset}>1.5$, while our CHAR mechanism makes $\text{IR}^{(\text{CHAR})}<1.05$.

\section{Conclusion}\label{section8}
In this paper, we focus on dynamic congestion games with endogenous time-varying conditions to analyze and regulate human-in-the-loop learning. In a typical parallel routing network, our analysis shows that the myopic routing policy (used by Google Maps and Waze) leads to severe under-exploration of stochastic paths. This results in a price of anarchy (PoA) greater than $2$, as compared to the social optimum achieved through optimal exploration-exploitation tradeoff. Besides, the myopic policy fails to ensure the correct learning convergence about users' traffic hazard beliefs. To address this, we first show that existing information-hiding mechanisms and deterministic path-recommendation mechanisms in Bayesian persuasion literature do not work with even \(\text{PoA}=\infty\). Accordingly, we propose a new combined hiding and probabilistic recommendation (CHAR) mechanism to hide all information from a selected user group and provide state-dependent probabilistic recommendations to the other user group. Our CHAR successfully ensures PoA less than \(\frac{5}{4}\), which cannot be further reduced by any other informational mechanism. Besides parallel networks, we further extend our analysis and CHAR to more general linear path graphs with multiple intermediate nodes, and we prove that the PoA results remain unchanged. Additionally, we carry out experiments with real-world datasets to validate the close-to-optimal performance of our CHAR mechanism.

In the future, we consider generalizing our human-in-the-loop learning beyond transportation applications, by considering stochastic queueing service systems (e.g., hospitals and restaurants). In queueing systems, many selfish users (e.g., customers for dining) are unwilling to explore new restaurants to learn useful information (e.g., tastes and queue length) for future customers. Thus, we need to design an efficient mechanism to regulate their exploration and exploitation of different servers.





\begin{IEEEbiography}[{\includegraphics[width=1in,height=1.25in,clip,keepaspectratio]{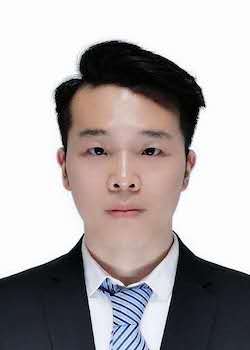}}]{Hongbo Li}
received the B.S. degree in Electronics and Electric Engineering from Shanghai Jiao Tong University, Shanghai, China, in 2019. He is currently working toward the Ph.D. degree with the Pillar of Engineering Systems and Design, Singapore University of Technology and Design (SUTD). His research interests include game theory and mechanism design, networked AI, and distributed learning.
\end{IEEEbiography}

\vspace{11pt}
\begin{IEEEbiography}[{\includegraphics[width=1in,height=1.25in,clip,keepaspectratio]{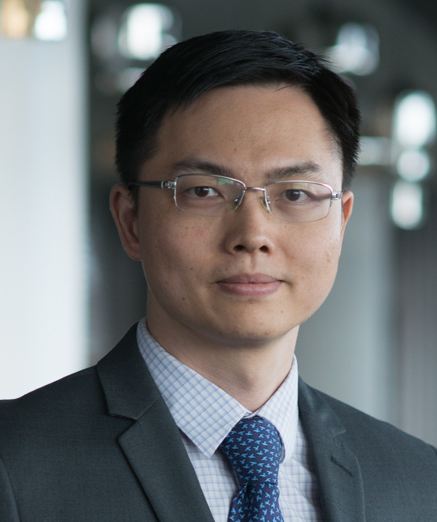}}]{Lingjie Duan}(Senior Member, IEEE) received the Ph.D. degree from The Chinese University of Hong Kong in 2012. He is an Associate Professor at the Singapore University of Technology and Design (SUTD) and is an Associate Head of Pillar (AHOP) of Engineering Systems and Design. In 2011, he was a Visiting Scholar at University of California at Berkeley, Berkeley, CA, USA. His research interests include network economics and game theory, cognitive communications, and cooperative networking.
He is an Associate Editor of IEEE Transactions on Mobile Computing and an Editor of IEEE/ACM Transactions on Networking. He was an Editor of IEEE Transactions on Wireless Communications and IEEE Communications Surveys and Tutorials. He also served as a Guest Editor of the IEEE Journal on Selected Areas in Communications Special Issue on Human-in-the-Loop Mobile Networks, as well as IEEE Wireless Communications Magazine. He received the SUTD Excellence in Research Award in 2016 and the 10th IEEE ComSoc Asia-Pacific Outstanding Young Researcher Award in~2015. He served as a General Chair of IEEE WiOpt~2023.
\end{IEEEbiography}

appendix

\subsection{Proof of Lemma 1}
Given $M=1$, a myopic user will compare the immediate travel costs of the two paths to choose the one that minimizes his own travel cost. Then there are three cases for the Nash equilibrium: 

If $c_1(0)\geq c_0(N(t))$, all the $N(t)$ users flock to safe path 0 with $n_1^{(m)}(t)=0$, which is the first case of (9). If $c_1(N(t))\leq c_0(0)$, all the $N(t)$ users flock to stochastic path~1 with $n_1^{(m)}(t)=N(t)$, which is the second case. Otherwise, the travel cost of each path is $c_0(N(t)-n_1^{(m)}(t))=c_1(n_1^{(m)}(t))$ in equilibrium. Solving this equation, we obtain the third case:
    \begin{align}
        {n_1^{(m)}(t)=}\frac{{N(t)}}{{2}}+\frac{{c_0(0)- c_1(0)}}{{2}}.\notag
    \end{align}

\subsection{Proof of Lemma 2}
Here we take the basic case $M=1$ with one stochastic path as an example to analytically prove the monotonicity. For the other cases with $M\geq 2$, the monotonicity still holds due to the homogeneity of stochastic paths. In the two-path network, if we can prove that $C^{(m)}\big(\mathbb{E}[\ell_2(t)|x'(t-1)],x(t),N(t)\big)$ increases with $\mathbb{E}[\ell_2(t)|x'(t-1)]$, $x(t)$ and $V(\cdot)$, then $C^*\big(\mathbb{E}[\ell_2(t)|x'(t-1)],x(t),N(t)\big)$ under the socially optimal policy also increases with them, as both long-term cost functions are the summation of discounted immediate costs $c(n^{(m)}(t),N(t))$ and $c(n^*(t),N(t))$ since time $t$.

We first prove the monotonicity of $C^{(m)}\big(\mathbb{E}[\ell_2(t)|x'(t-1)],x(t),N(t)\big)$ with respect to travel latency $\mathbb{E}[\ell_2(t)|x'(t-1)]$. Suppose that at current time $t$, there are two initial latencies satisfying $\mathbb{E}[\ell_a|x'(t-1)]<\mathbb{E}[\ell_b|x'(t-1)]$ of stochastic path 2. Accordingly, the corresponding myopic exploration numbers satisfy $n_a^{(m)}(t)\leq n_b^{(m)}(t)$ (which we will prove later in Appendix C). Then we compare the two cost functions under the two initial travel latencies:
\begin{align}
    &C^{(m)}\big(\mathbb{E}[\ell_a|x'(t-1)],x(t),N(t)\big)-\notag\\&C^{(m)}\big(\mathbb{E}[\ell_b|x'(t-1)],x(t),N(t)\big)\notag\\ \leq & n^{(m)}_b(t)\Big(\mathbb{E}[\ell_a|x'(t-1)]-\mathbb{E}[\ell_b|x'(t-1)]\Big)\notag\\&+\rho \Big(Q^{(m)}_a(t+1,n_b^{(m)}(t))-Q^{(m)}_b(t+1,n_b^{(m)}(t))\Big)\notag\\
    < & \rho \Big(Q^{(m)}_a(t+1,n_b^{(m)}(t))-Q^{(m)}_b(t+1,n_b^{(m)}(t))\Big),\label{20}\tag{23}
\end{align}
where the first inequality is derived by taking $n_a(t)=n^{(m)}_b(t)$, as this exploration number $n^{(m)}_b(t)$ enlarges the immediate cost of $C\big(\mathbb{E}[\ell_a|x'(t-1)],x(t),N(t)\big)$ as compared to $C^{(m)}\big(\mathbb{E}[\ell_a|x'(t-1)],x(t),N(t)\big)$ under $n_a^{(m)}(t)$. We can use the same method to further expand $Q^{(m)}_a(t+1,n_b^{(m)}(t))$ and $Q^{(m)}_b(t+1,n_b^{(m)}(t))$ in (\ref{20}) for the next time slot $t$ to obtain 
\begin{align*}
    &\rho \Big(Q^{(m)}_a(t+1,n_b^{(m)}(t))-Q^{(m)}_b(t+1,n_b^{(m)}(t))\Big)\\ <& \rho^2 \Big(Q^{(m)}_a(t+2,n_b^{(m)}(t+1))-Q^{(m)}_b(t+2,n_b^{(m)}(t+1))\Big)\\
    <& \cdots< 0,
\end{align*}
as the immediate cost under travel latency $\mathbb{E}[\ell_a|x'(t-1)]$ is less than that under $\mathbb{E}[\ell_b|x'(t-1)]$ per time slot since time $t$.
Therefore, the cost function under the myopic policy increases with $\mathbb{E}[\ell_2(t)|x'(t-1)]$.  

Next, we consider two different belief states $x_a$ and $x_b$ with $x_a<x_b$ to make a comparison between their caused cost functions. Suppose $\mathbb{E}[\ell(t)|x_a]=\mathbb{E}[\ell(t)|x_b]$ at current time $t$. We obtain
\begin{align*}
    &C^{(m)}\big(\mathbb{E}[\ell(t)|x_a],x_a,N(t)\big)-C^{(m)}\big(\mathbb{E}[\ell(t)|x_b],x_b,N(t)\big)\\ \leq & \rho \Big(Q^{(m)}_a(t+1,n_b^{(m)}(t))-Q^{(m)}_b(t+1,n_b^{(m)}(t))\Big)<0,
\end{align*}
where the first inequality is similarly derived by setting $n_a(t)=n^{(m)}_b(t)$ as in (\ref{20}), and the second inequality is because of $\mathbb{E}[\ell(t+1)|x'_a(t)]<\mathbb{E}[\ell(t+1)|x'_b(t)]$ in the cost-to-go, given $x_b>x_a$. Hence, the cost function under the myopic policy increases with hazard belief $x(t)$. 

Finally, it is obvious that both cost functions increase with $V(\cdot)$, as $V(\cdot)$ is a constant for immediate cost per time slot. Therefore, we skip the detailed proof here, and this completes the proof of cost functions' monotonicity.

\subsection{Proof of Lemma 3}
We first prove the monotonicity of exploration numbers $n^{(m)}_i(t)$ and $n^*_i(t)$ with respect to hazard belief $x_i(t)$ and error cost $V(\cdot)$. Based on the results, we then prove the monotonicity of $n^{(m)}_i(t)-n^*_i(t)$. Here we take the basic case $M=1$ with one stochastic path as an example to analytically prove the monotonicity. For the other cases with $M\geq 2$, the monotonicity still holds due to the homogeneity of stochastic paths. 

\subsubsection{Monotonicity of Exploration Numbers}
We first discuss the monotonicity of $n_1^{(m)}(t)$ based on its definition in (10). 
By combining(10) with (8) and (9), we rewrite $n_1^{(m)}(t)$ to
\begin{align}\label{22}
    \frac{\Scale[0.92]{N(t)+\ell_0-\mathbb{E}[\ell_1(t)|x_1'(t-1)]-V(n_1^{(m)}(t-1))}}{2}.\tag{24}
\end{align}
It is obvious that $n_1^{(m)}(t)$ linearly decreases with $\mathbb{E}[\ell_1(t)|x_1'(t-1)]$ and $V(\cdot)$. Therefore, given expected latency $\mathbb{E}[\ell_1(t-1)|x'_1(t-2)]$ at the last time slot, if hazard belief $x_1(t)$ increases, the current $\mathbb{E}[\ell_1(t)|x_1'(t-1)=x_1(t)]$ also becomes longer, which consequently reduces exploration number $n_1^{(m)}(t)$.

Then we further prove the monotonicity of $n_1^*(t)$ by solving the long-term objective function under the socially optimal policy in (13). Here we focus on the general case with $n_1^*(t)\geq 1$ to calculate $n_1^*(t)$. As $n_1^*(t)$ is derived at the extreme point of the cost function $C^*\big(\mathbb{E}[\ell_1(t)|x_1'(t-1)],x_1(t),N(t)\big)$, we need to solve the first-order-derivative condition $\frac{\partial C^*\big(\mathbb{E}[\ell_1(t)|x_1'(t-1)],x_1(t),N(t)\big)}{\partial n_1^*(t)}=0$. Then we obtain
\begin{align}
    n_1^*(t)=&\frac{N(t)}{2}+\frac{\ell_0-\mathbb{E}[\ell_1(t)|x_1'(t-1)]-V(n_1^*(t-1))}{4}\notag\\-&\frac{\rho}{4}\frac{\partial \mathbb{E}[C^*(\mathbb{E}[\ell_1(t+1)|x_1'(t)],x_1(t+1),\Bar{N})]}{\partial n_1^*(t)}.\label{23}\tag{25}
\end{align}
Next, we prove $n_1^*(t)$'s decreases with $V(n_1^*(t-1))$ by:
\begin{align*}
    \frac{\partial n^*(t)}{\partial V(n^*(t-1))}
    =&-\frac{1}{4}-\frac{\rho}{4}\frac{\partial\frac{\partial \mathbb{E}[C^*(\mathbb{E}[\ell_1(t+1)|x_1'(t)],x_1(t+1),\Bar{N})]}{\partial n^*(t)}}{\partial V(n_1^*(t-1))},
\end{align*}
which is less than $0$ due to $\frac{\partial\frac{\partial \mathbb{E}[C^*(\mathbb{E}[\ell_1(t+1)|x_1'(t)],x_1(t+1),\Bar{N})]}{\partial n^*(t)}}{\partial V(n_1^*(t-1))}=0$, as current observation error $V(n_1^*(t-1))$ has no effect on cost-to-go since the next time slot. Based on the above analysis, we obtain that $n_1^*(t)$ decreases with $V(\cdot)$. Then we can use the same method to prove $\frac{\partial n_1^*(t)}{\partial x_1(t)}\leq 0$ to show that $n_1^*(t)$ also decreases with hazard belief $x_1(t)$. 

\subsubsection{Monotonicity of $n^{(m)}_i(t)-n^*_i(t)$}
Based on the monotonicity of $n^{(m)}_i(t)$ and $n^*_i(t)$ above, we will prove $n_i^*(t)-n_i^{(m)}(t)$ increases with $x_i(t)$ by showing $\frac{\partial (n_i^*(t)-n_i^{(m)}(t))}{\partial x_i(t)}\geq 0$. If $M=1$, according to (\ref{22}) and (\ref{23}), we obtain
\begin{align}\label{24}
    &\frac{\partial (n_1^*(t)-n_1^{(m)}(t))}{\partial x_1(t)}\notag\\=&-\frac{1}{4}\frac{\mathbb{E}[\ell_1(t)|x_1'(t-1)]}{\partial x_1(t)}+\frac{1}{2}\frac{\mathbb{E}[\ell_1(t)|x_1'(t-1)]}{\partial x_1(t)}\notag\\&-\frac{\rho}{4}\frac{\partial^2 \mathbb{E}[C^*(\mathbb{E}[\ell_1(t+1)|x_1'(t)],x_1(t+1),\Bar{N})]}{\partial n_1^*(t)\partial x_1(t)}\notag\\ =&\frac{1}{4}\frac{\mathbb{E}[\ell_1(t)|x_1'(t-1)]}{\partial x_1(t)}\notag\\ &-\frac{\rho}{4}\frac{\partial^2 \mathbb{E}[C^*(\mathbb{E}[\ell_1(t+1)|x_1'(t)],x_1(t+1),\Bar{N})]}{\partial n_1^*(t)\partial x_1(t)}.\tag{26}
\end{align}
Based on the socially optimal policy in (13), we expand $\mathbb{E}[C^*(\mathbb{E}[\ell_1(t+1)|x_1'(t)],x_1(t+1),\Bar{N})]$ since $t+1$ as:
\begin{align*}
    &\mathbb{E}[C^*(\mathbb{E}[\ell_1(t+1)|x_1'(t)],x_1(t+1),\Bar{N})]\\=&\mathbf{Pr}(y_1(t)=1|n^*_1(t))C^*(\mathbb{E}[\ell_1(t+1)|x_1'(t),y_1(t)=1])\\&+\mathbf{Pr}(y_1(t)=0|n^*_1(t))C^*(\mathbb{E}[\ell_1(t+1)|x_1'(t),y_1(t)=0])
\end{align*}
Then we calculate $\frac{\partial^2 \mathbb{E}[C^*(\mathbb{E}[\ell_1(t+1)|x_1'(t)],x_1(t+1),\Bar{N})]}{\partial n_1^*(t)\partial x_1(t)}$ in (\ref{24}) as:
\begin{align*}
     &\frac{\partial^2\mathbb{E}[C^*(\mathbb{E}[\ell_1(t+1)|x_1'(t)],x_1(t+1),\Bar{N})]}{\partial n_1^*(t)\partial x_1(t)}\\=&\frac{\partial^2 \mathbf{Pr}(y_1(t)=1|n_1^*(t))}{\partial n_1^*(t)\partial x_1(t)}C^*(\mathbb{E}[\ell_1(t+1|x_1'(t),1)])\\ &+\frac{\partial C^*(\mathbb{E}[\ell_1(t+1|x_1'(t),1)])}{\partial x_1(t)}\frac{\partial \mathbf{Pr}(y_1(t)=1|n_1^*(t))}{\partial n_1^*(t)}\\ &+\frac{\partial^2 \mathbf{Pr}(y_1(t)=0|n_1^*(t))}{\partial n_1^*(t) \partial x_1(t)} C^*(\mathbb{E}[\ell_1(t+1|x_1'(t),0)])\\ &+\frac{\partial C^*(\mathbb{E}[\ell_1(t+1|x_1'(t),0)])}{\partial x_1(t)}\frac{\partial \mathbf{Pr}(y_1(t)=0|n_1^*(t))}{\partial n_1^*(t)}.
\end{align*}
Taking the above equation back to (\ref{24}), we obtain
\begin{align*}
    &\frac{\partial (n_1^*(t)-n_1^{(m)}(t))}{\partial x_1(t)}\\ \geq& \frac{1}{4}\frac{\mathbb{E}[\ell_1(t)|x_1'(t-1)]}{\partial x_1(t)}\\&-\frac{\rho}{4}\frac{\partial C^*(\mathbb{E}[\ell_1(t+1|x_1'(t),1)])}{\partial x_1(t)}\frac{\partial \mathbf{Pr}(y_1(t)=1|n_1^*(t))}{\partial n_1^*(t)}\\ &+\frac{\rho}{4}\frac{\partial C^*(\mathbb{E}[\ell_1(t+1|x_1'(t),0)])}{\partial x_1(t)}\frac{\partial \mathbf{Pr}(y_1(t)=1|n_1^*(t))}{\partial n_1^*(t)}
\end{align*}
\begin{align*}
    \geq& \frac{1}{4}\frac{\mathbb{E}[\ell_1(t)|x_1'(t-1)]}{\partial x_1(t)}\\&-\frac{\rho}{4}\frac{\partial C^*(\mathbb{E}[\ell_1(t+1|x_1'(t),1)])}{\partial x_1(t)}\frac{\partial \mathbf{Pr}(y_1(t)=1|n_1^*(t))}{\partial n_1^*(t)}\\ &+\frac{\rho}{4}\frac{\partial C^*(\mathbb{E}[\ell_1(t+1|x_1'(t),1)])}{\partial x_1(t)}\frac{\partial \mathbf{Pr}(y_1(t)=1|n_1^*(t))}{\partial n_1^*(t)}\\ =&\frac{1}{4}\frac{\mathbb{E}[\ell_1(t)|x_1'(t-1)]}{\partial x_1(t)}>0,
\end{align*}
where the first inequality is because of $\mathbf{Pr}(y_1(t)=0|n_1^*(t))=1-\mathbf{Pr}(y_1(t)=1|n_1^*(t))$, and the last inequality is due to that $\mathbb{E}[\ell_1(t)|x_1'(t-1)]$ increases with $x_1(t)$. In summary, the difference $n_1^*(t)-n_1^{(m)}(t)$ increases with $x_1(t)$. 
 
Next, we use the same method to prove that $(n_1^*(t)-n_1^{(m)}(t))$ decreases with observation error function $V(\cdot)$:
\begin{align*}
    &\frac{\partial (n_1^*(t)-n_1^{(m)}(t))}{\partial V(n_1^*(t-1))}\\=&\frac{1}{4}-\frac{\rho}{4}\frac{\partial^2 \mathbb{E}[C^*(\mathbb{E}[\ell_1(t+1)|x_1'(t)],x_1(t+1),\Bar{N})]}{\partial n^*(t)\partial V(n^*(t-1))},
\end{align*}
which is larger than $0$ as $\frac{\partial^2 \mathbb{E}[C^*(\mathbb{E}[\ell_1(t+1)|x_1'(t)],x_1(t+1),\Bar{N})]}{\partial n_1^*(t)\partial V(n_1^*(t-1))}=0$. This completes the proof.

\subsection{Proof of Proposition 1}
Given $\mathbb{E}[\ell_i(t-1)|x_i'(t-2)]$ on stochastic path $i$ at $t-1$, we first prove that there exists a high belief $x_i(t)=R$ to make the myopic policy under-explore path $i$ (with $n_i^*(t)-n_i^{(m)}(t)>0$). Then we prove that there exists another low belief $x_i(t)=r<R$ to make the myopic policy over-explore (with $n_i^*(t)-n_i^{(m)}(t)\leq 0$). Based on Lemma~2 that $n_i^*(t)-n_i^{(m)}(t)$ increases with $x_i(t)$, there exists an unique exploration threshold $x_{th}\in(r,R)$, which also increases with $V(\cdot)$. 

\subsubsection{Proof of Under-exploration}\label{under-explore}
Let hazard belief $x_i(t)=R$ satisfies $\mathbb{E}[\alpha_i(t)|R]=1$ for any stochastic path $i\in\{1,\cdots,M\}$. If $\ell_0+N(t)\leq\mathbb{E}[\ell_i(t)|R]+V(0)$, all the $N(t)$ users under the myopic policy will choose safe path~0. Then the caused long-term expected social cost is
\begin{align*}
    C^{(m)}(\mathbf{L}(t),\mathbf{x}(t),N(t))=\sum_{t=0}^{\infty}\rho N(\ell_0+\Bar{N})=\frac{\Bar{N}(\ell_0+\Bar{N})}{1-\rho}.
\end{align*}
However, the socially optimal policy may still recommend $n^*_i(t)$ users to explore a stochastic path $i$ to obtain 
\begin{align}\label{25}
    &C^*(\mathbf{L}(t),\mathbf{x}(t),N(t)) \notag\\ \leq &n_i^*(t)\big(\mathbb{E}[\ell_i(t)|R]+n_i^*(t)+V(0)\big)+(N(t)-n_i^*(t))\ell_0 \notag\\ &+\rho \mathbb{E}_{y_i(t)}[C^*(\mathbf{L}(t+1),\mathbf{x}(t+1),\Bar{N})].\tag{27}
\end{align}

In (\ref{25}), there are two cases to update cost-to-go $\mathbb{E}_{y_i(t)}[C^*(\mathbf{L}(t+1),\mathbf{x}(t+1),\Bar{N})]$ in future time slots.
If $y_i(t)=1$, for any future user arrivals at $\tau>t$, the socially optimal policy will recommend $n_i^*(\tau)=0$ users to explore stochastic path $i$ to avoid extra travel cost. Thus,
\begin{align*}
    \Scale[0.94]{C^*(\mathbf{L}(t+1),\mathbf{x}(t+1),\Bar{N}|y_i(t)=1)}\leq& \frac{\Bar{N}(\ell_0+\Bar{N})}{1-\rho}\\=&\Scale[0.94]{C^{(m)}(\mathbf{L}(t),\mathbf{x}(t),\Bar{N})}.
\end{align*}
While if $y_i(t)=0$, the expected travel latency for $t+1$ becomes $0$ due to low-hazard state $\alpha_L=0$. As the cost function increases with travel latency on path $i$, the cost-to-go satisfies
\begin{align*}
    C^*(\mathbf{L}(t+1),\mathbf{x}(t+1),\Bar{N}|y_i(t)=0)<&C^*(\mathbf{L}(t),\mathbf{x}(t),\Bar{N})\\\leq &C^{(m)}\big(\mathbf{L}(t),\mathbf{x}(t),\Bar{N}\big).
\end{align*}
In summary, the cost-to-go under socially optimal policy is always smaller than that under the myopic policy.
If $\ell_0\gg 0$, $N(t)\ll \ell_0$ and $V(N(t))\ll\ell_0$, we further calculate (\ref{25}):
\begin{align*}
    &C^*(\mathbf{L}(t),\mathbf{x}(t),N(t))
    \\\leq &n_i^*(t)\big(\mathbb{E}[\ell_i(t)|R]+n_i^*(t)+V(0)\big)+(N(t)-n_i^*(t))\ell_0\\ &+\rho \mathbf{Pr}(y(t)=0)C^*(\mathbf{L}(t+1),\mathbf{x}(t+1),\Bar{N}|y_i(t)=0)\\ &+\rho \mathbf{Pr}(y(t)=1)C^*(\mathbf{L}(t+1),\mathbf{x}(t+1),\Bar{N}|y_i(t)=1)\\
    <&N(t)\ell_0+\rho C^{(m)}(\mathbf{L}(t),\mathbf{x}(t),\Bar{N})
   \\ =&C^{(m)}(\mathbf{L}(t),\mathbf{x}(t),N(t)).
\end{align*}
Hence, if $x_i(t)\geq R$, current $N(t)$ users under-explore stochastic path $i$ under the myopic policy.

\subsubsection{Proof of Over-exploration}\label{over-explore}
Next, we prove that there exists a smaller belief $x_i(t)=r<R$ leading to the myopic policy's over-exploration on path $i$.
Based on (10), if $\ell_0>\mathbb{E}[\ell_i(t)|r]+N(t)+V(N(t-1))$, all the $N(t)$ myopic users choose to explore stochastic path $i$, leading to $n_i^*(t)\leq n_i^{(m)}(t)$ (over-exploration). However, we will still try to prove there exist parameters to strictly make $n_i^*(t)< n_i^{(m)}(t)$.

Given the above $\mathbf{L}(t),\mathbf{x}(t)$ and $N(t)$, we calculate:
\begin{align*}
    \Scale[0.93]{C^{(m)}(\mathbf{L}(t),\mathbf{x}(t),N(t))=}&\Scale[0.93]{N(t)(\mathbb{E}[\ell_i(t)|r]+N(t)+V(N(t-1))}\\&\Scale[0.93]{+ \rho \mathbb{E}_{y_i(t)}[C^{(m)}(\mathbf{L}(t+1),\mathbf{x}(t+1),\Bar{N})]},
\end{align*}
where the cost-to-go since next time slot $t+1$ satisfies
\begin{align}
    &\mathbb{E}_{y_i(t)}[C^{(m)}(\mathbf{L}(t+1),\mathbf{x}(t+1),\Bar{N})]\tag{28}\label{26}\\=&\Scale[0.95]{\mathbf{Pr}(y_i(t)=0|N(t)) C^{(m)}(\mathbf{L}(t+1),\mathbf{x}(t+1),\Bar{N}|y_i(t)=0)}\notag\\
    &\Scale[0.95]{+\mathbf{Pr}(y_i(t)=1|N(t)) C^{(m)}(\mathbf{L}(t+1),\mathbf{x}(t+1),\Bar{N}|y_i(t)=1)}.\notag
\end{align}

We first assume that the optimal exploration number $n^*_i(t)< N(t)$. Then, if the group observation probabilities satisfy $q_L(N(t))>0, q_H(N(t))=1$ and $\alpha_L=0$, we obtain
\begin{align*}
    \mathbf{Pr}(y_i(t)=1|N(t))=&rq_H(N(t))+(1-r)q_L(N(t))\\=&r+(1-r)q_L(N(t))\\>&\mathbf{Pr}(y_i(t)=1|n_i^*(t)),
\end{align*}
due to the fact that $q_L(n_i(t))$ decreases with $n_i(t)$. Similarly, the posterior belief under $y_i(t)=1$ satisfies
\begin{align*}
    x'_i(t|y_i(t)=1,N(t))=&\frac{rq_H(N(t))}{rq_H(N(t))+(1-r)q_L(N(t))}\\ >&x_i'(t|y_i(t)=1,n_i^*(t)),
\end{align*}
as $q_H(n(t))$ increases with $n(t)$. 

Then the cost-to-go under the myopic policy in (\ref{26}) satisfies
\begin{align*}
    &\mathbb{E}_{y_i(t)}[C^{(m)}(\mathbf{L}(t+1),\mathbf{x}(t+1),\Bar{N})]\\=&\Scale[0.95]{\mathbf{Pr}(y_i(t)=0|N(t)) C^{(m)}(\mathbf{L}(t+1),\mathbf{x}(t+1),\Bar{N}|y_i(t)=0)}\\ 
    &+\Scale[0.95]{\mathbf{Pr}(y_i(t)=1|N(t)) C^{(m)}(\mathbf{L}(t+1),\mathbf{x}(t+1),\Bar{N}|y_i(t)=1)}\\
    >&\Scale[0.95]{\mathbf{Pr}(y_i(t)=0|n^*_i(t)) C^{(m)}(\mathbf{L}(t+1),\mathbf{x}(t+1),\Bar{N}|y_i(t)=0)}\\
    &\Scale[0.95]{+\mathbf{Pr}(y_i(t)=1|n^*_i(t)) C^{(m)}(\mathbf{L}(t+1),\mathbf{x}(t+1),\Bar{N}|y_i(t)=1)}\\
    =&\mathbb{E}_{y_i(t)}[C^*(\mathbf{L}(t+1),\mathbf{x}(t+1),\Bar{N})].
\end{align*}
As the cost-to-go $\mathbb{E}_{y_i(t)}[C^{(m)}(\mathbf{L}(t+1),\mathbf{x}(t+1),\Bar{N})]$ under $n^{(m)}_i(t)=N(t)$ is greater than $\mathbb{E}_{y_i(t)}[C^*(\mathbf{L}(t+1),\mathbf{x}(t+1),\Bar{N})]$ under $n^*_i(t)<N(t)$, current $N(t)$ users under the myopic policy over-explore stochastic path $i$ if $x_i(t)\leq r<R$.

In summary, we have proved $n_i^*(t)> n_i^{(m)}(t)$ for $x_i(t)\geq R$ in Appendix \ref{under-explore} and $n_i^*(t)\leq n_i^{(m)}(t)$ for $x_i(t)\leq r$ in Appendix \ref{over-explore}. Based on Lemma 3 that $n_i^*(t)-n_i^{(m)}(t)$ increases with hazard belief $x_2(t)$, there must exist unique exploration threshold $x_{th}\in (r,R)$, which increases with $V(\cdot)$.

\subsection{Proof of Theorem 1}

\begin{figure*}
	\begin{subequations}
		\begin{align}
		      \text{PoA}^{(m)}\tag{29}\label{27}\geq& \frac{\frac{1-\rho^k}{1-\rho}(\ell_0(t)+N(t))\cdot N(t)}{N(t)\cdot(\mathbb{E}[\ell_i(t)|x_i'(t-1)]+N(t)+V(N(t)))+\sum_{j=1}^k(\frac{(\rho \alpha_H)^{j-1}-1}{\rho\alpha_H-1}\rho \alpha_H+N(t)+V(N(t)))\cdot N(t)}\\
            \geq &\frac{\frac{1-\rho^k}{1-\rho}N(t)\ell_0(t)}{N(t)\ell_0(t)+\frac{\ell_0(t)}{2}N(t)\frac{\rho-\rho^k}{1-\rho}}
            =\lim_{k\rightarrow \infty}\frac{2(1-\rho^k)}{2-\rho-\rho^k}=2.\notag
		\end{align}
	\end{subequations}
	{\noindent} \rule[-10pt]{18cm}{0.05em}
\end{figure*}

We prove this theorem by analyzing the worst-case scenario with the myopic policy's zero-exploration. However, the socially optimal policy recommends some users explore path $i$ to find possible $\alpha_L$ and reduce travel costs for future users. 

Initially, we set $\ell_0=\mathbb{E}[\ell_i(t)|x_i(t)]+V(N(t))$, $\alpha_L=0$ and $\alpha_H>1$ with $\mathbb{E}[\alpha_i(t)|x_i(t)]=1$, such that myopic users will never explore any stochastic path $i$ to avoid long travel latency there. As there is no information update and $\mathbb{E}[\alpha_i(t)|x_i(t)]=1$, the expected travel latency on stochastic path $i$ remains unchanged. Then the caused long-term expected social cost is
\begin{align*}
    C^{(m)}(\mathbf{L}(t),\mathbf{x}(t),N(t))=\frac{\Bar{N}(\ell_0+\Bar{N})}{1-\rho}.
\end{align*}

For the socially optimal policy, it lets $\frac{N(t)}{M}$ users explore each stochastic path $i$ to derive immediate social cost:
\begin{align*}
    c_i(n_i^*(t))=\frac{N(t)}{M}(\mathbb{E}[\ell_i(t)|x_i'(t-1)]+\frac{N(t)}{M}+V(0)).
\end{align*}

Suppose that the system has been running for a long time $t$ before the current time slot. Then the probability of the actual travel latency on stochastic path $i$ being reduced to $\mathbb{E}[\ell_i(t)]=0$ by low hazard state $\alpha_L=0$ is $\mathbf{Pr}(\mathbb{E}[\ell_i(t)]=0)=1-x_i(t)^t\rightarrow 0$.
Accordingly, it is almost sure for current $N(t)$ users to observe $\alpha_i(t)=\alpha_L$ on each path $i$. 
After that, the travel cost $c_i(\frac{N_{min}}{M})$ of path $i$ gradually increases to $c_0(0)$ again at after $k$ time slots. In the worst case, all the $N(t)=\frac{N_{max}}{M}$ users always observe $y_i(t)=1$ to make $\alpha(t)=\alpha_H$. Based on the linear correlation function of $\mathbb{E}[\ell_i(t+1)|x_i'(t)]$, the travel latency for $(t+k)$-th time slot satisfies
\begin{align*}
    \mathbb{E}[\ell_i(t+k)]\leq \sum_{j=1}^k \alpha_H^j \frac{N_{max}}{M}=\frac{\alpha_H^{k-1}-1}{\alpha_H-1}\alpha_H \frac{N_{max}}{M}.
\end{align*}
Based on this inequality, we solve $c_i(\frac{N_{min}}{M})=c_0(0)$ to obtain
\begin{align*}
    k\geq 1+\log_{\alpha_H}\left(M\frac{(\ell_1-\frac{N_{min}}{M}-V(\frac{N_{min}}{M})(\alpha_H-1)}{\alpha_H N_{max}}+1\right).
\end{align*}
It means that users' travel costs on the two paths become the same again after $k$ slots. Based on our analysis above, we calculate PoA in (\ref{27}), where the second inequality is due to the convexity in $k$ of the second term in the denominator.

\subsection{Proof of Lemma 4}
We prove this lemma by analyzing the same worst-case as Theorem 1 here.

At the beginning of time $t$, if $\mathbb{E}[\alpha_i(t)|x_i'(t-1)]\geq 1$ and $\mathbb{E}[\ell_i|x_i'(t-1)]\geq \ell_0-V(n_i(t-1))$, selfish myopic users will never explore stochastic path $i$. In this case, $x_i(t)$ remains unchanged as $x_i'(t-1)$ and will never converge to $\Bar{x}$.

Therefore, according to the worst-case above, the myopic policy cannot ensure $x(t)$ to correctly converge to its actual value $\Bar{x}$.

\subsection{Proof of Proposition 2}

For the socially optimal policy, we will prove that if $x_i(t)>\Bar{x}$, its updated $x_i(t+1)$ at $t+1$ is expected to decrease, i.e., $x_i(t+1)<x_i(t)$. While if $x_i(t)<\Bar{x}$, $x_i(t+1)$ is expected to be greater than $x_i(t)$. Based on the two cases, we obtain that $x_i(t)$ will finally converge to the real steady state $\Bar{x}$ under consequent explorations of stochastic paths.

If the actual hazard belief $x_i(t)>\Bar{x}$, then users' expected probability of observing a hazard is
\begin{align*}
    \Scale[0.93]{\mathbf{Pr}(y_i(t)=1|n(t),x_i(t))=}&\Scale[0.93]{(1-x_i(t))q_L(n_i(t))+x_i(t)q_H(n_i(t))}\\
    >& \Scale[0.93]{(1-\Bar{x})q_L(n_i(t))+\Bar{x}q_H(n_i(t))}\\
    =&\Scale[0.93]{\mathbf{Pr}(y_i(t)=1|n_i(t),\Bar{x})},
\end{align*}
based on the fact that $q_H(n_i(t))>q_L(n_i(t))$ and $x_i(t)>\Bar{x}$. It means the actual probability $\mathbf{Pr}(y_i(t)=1|n_i(t),\Bar{x})$ for current $n_i(t)$ users to observe a hazard ($y_i(t)=1$) under $\Bar{x}$ is lower than the expected probability $\mathbf{Pr}(y_i(t)=1|n_i(t),x_i(t))$. Similarly, we obtain $\mathbf{Pr}(y_i(t)=0|n_i(t),x_i(t)) <\mathbf{Pr}(y_i(t)=0|n_i(t),\Bar{x})$,
i.e., the actual probability to observe $y_i(t)=0$ under $\Bar{x}$ is greater than the expected probability under $x_i(t)$.

Based on the above analysis, if there are users traveling on stochastic path $i$ and sharing their observation summary $y_i(t)$, the actually updated belief at the next time slot $t+1$ satisfies
\begin{align*}
    &x_i(t+1)\\=&\mathbf{Pr}(y_i(t)=1|\Bar{x})x'_i(t|1)+\mathbf{Pr}(y_i(t)=0|\Bar{x})x'_i(t|0)\\
    <& \mathbf{Pr}(y_i(t)=1|x_i(t))x'_i(t|1)+\mathbf{Pr}(y_i(t)=0|x_i(t))x'_i(t|0)\\ =& \mathbb{E}_{y_i(t)}[x_i(t+1)|x_i(t)] = x_i(t),
\end{align*}
which means that the hazard belief $x_i(t)$ will actually decreases to $x_i(t+1)$ if $x_i(t)>\Bar{x}$. Similarly, we can prove that the actually updated belief $x_i(t+1)> x_i(t)$ if $x_i(t)<\Bar{x}$. 

In summary, under the socially optimal policy with users' consequent exploration and learning on stochastic path $i\in\{1,\cdots,M\}$, $x_i(t)$ can finally converge to $\Bar{x}$ as $t\rightarrow \infty$.

\subsection{Proof of Lemma 5}\label{Proof_G}

Under the information hiding mechanism, users can only infer that there are $\mathbb{E}[N(t)]=\Bar{N}$ users arriving currently. 

Initially, let $\ell_0\geq \mathbb{E}_{\Bar{x}\sim \mathbb{P}(\Bar{x})}[\Bar{\ell_i}|\Bar{x}]+V(0)+\frac{N_{max}}{M}$, such that $n_i^{\emptyset}=N(t)$ in (16), as $c_0(0)-\mathbb{E}_{\Bar{x}\sim \mathbb{P}(\Bar{x})}[c_i(0)|\Bar{x}]\geq \frac{N_{max}}{M}$. At the same time, we set actual $\mathbb{E}[\ell_i(0)|x_i(0)]\gg \ell_0$ and $\mathbb{E}[\alpha_i(0)|x_i(0)]>1$. In this case, the social optimum will recommend all the $N(t)$ users travel safe path 0 to reduce the latency on path $i$. The caused PoA under over-exploration is:
\begin{align*}
    \text{PoA}\geq \frac{\sum_{j=0}^\infty \rho^j \Bar{N}(\ell_i(0)+\frac{\Bar{N}}{M}+V(\frac{\Bar{N}}{M}))}{\sum_{j=0}^\infty \rho^j \Bar{N}\ell_0} =\infty.
\end{align*}

While if $c_0(N_{max})-\mathbb{E}_{\Bar{x}\sim \mathbb{P}(\Bar{x})}[c_i(0)|\Bar{x}]\leq -N_{max}$ and the initial actual travel latency on stochastic path $i$ satisfies $\mathbb{E}[\ell_i(0)|x_i(0)]\ll \ell_0$, then the information-hiding mechanism leads to under-exploration of stochastic path $i$.

\subsection{Proof of Lemma 6}

We prove that the recommendation-only mechanism makes PoA infinite by showing that, if $\Bar{N}\gg 1$, users will still follow $n_i^{\emptyset}$ under the information hiding policy in (16). 

Under the recommendation-only mechanism, if a user is recommended to choose stochastic path $\pi(t)=i$, his posterior distribution of the long-run expected belief $\Bar{x}$ changes from $\mathbb{P}(\Bar{x})$ to $\mathbb{P}'(\Bar{x}|n_i^*(t)\geq 1)$. Then he estimates the expected travel latency on stochastic path $i$ as $\mathbb{E}_{\Bar{x}\sim \mathbb{P}'(\Bar{x}|n_i^*(t)\geq 1)} [c_i(1)|\Bar{x}]\approx 
    \mathbb{E}_{\Bar{x}\sim \mathbb{P}(\Bar{x})} [c_i(1)|\Bar{x}],$
given $\Bar{N}\gg 1$. In consequence, this user will not follow this recommendation $\pi(t)=i$ if $\mathbb{E}_{\Bar{x}\sim \mathbb{P}(\Bar{x})} [c_i(1)|\Bar{x}]>c_0(N-1)$, even with all the other $\Bar{N}-1$ users choosing path 0. 
Similarly, if a user is recommended to choose safe path $\pi(t)=0$, his posterior distribution becomes $\mathbb{P}'(\Bar{x}|n_i^*(t)\geq 0)$, which always equals $\mathbb{P}(\Bar{x})$ without any other information. Therefore, each user still follows $n_i^{\emptyset}$ in (16) to make his path decision, leading to $\text{PoA}=\infty$ as in Lemma 5.

\subsection{Proof of Theorem 2}
We first prove that each user in the recommendation-only group is incentive compatible to follow recommendation $\pi(t)$, and there always exists $p_L$ and $p_H$ to satisfy the condition $p_L\mathbb{P}(x_{th})\geq p_H(1-\mathbb{P}(x_{th}))$. Then we prove that by dynamically solving (17) to decide number $N^{\emptyset}(t)$, our CHAR realizes the PoA in (21). Finally, we show that this PoA in (21) is the minimum achievable and cannot be reduced by any other informational mechanism. 

\subsubsection{Incentive Compatibility for all Users}
If a user of the recommendation-only group is recommended $\pi(t)=i$, he estimates that there will be $\Bar{n}_i^{(\text{CHAR})}$ users on path~$i$ under CHAR. Then his posterior probability of $x_i(t)<x_{th}$ becomes:
\begin{align*}
    &\mathbf{Pr}(x_i(t)<x_{th}|\pi(t)=i)\\=& \frac{\mathbf{Pr}(\pi(t)=i,x_i(t)<x_{th})}{\mathbf{Pr}(\pi(t)=i)}
    \\=&\frac{\mathbf{Pr}(\pi(t)=i|x_i(t)<x_{th})\mathbf{Pr}(x_i(t)<x_{th})}{\mathbf{Pr}(\pi(t)=i)}\\=&\frac{\mathbb{P}(x_{th})p_L}{\mathbb{P}(x_{th})p_{L}+(1-\mathbb{P}(x_{th}))p_H},
\end{align*}
where the last equality is derived by replacing $\mathbf{Pr}(\pi(t)=i)$ with $\mathbf{Pr}(\pi(t)=i|x_i(t)<x_{th})\mathbf{Pr}(x_i(t)<x_{th})+\mathbf{Pr}(\pi(t)=i|x_i(t)\geq x_{th})\mathbf{Pr}(x_i(t)\geq x_{th})$ in the second equality, according to the law of total probability.
Similarly, his estimated posterior probability $\mathbf{Pr}(x_i(t)\geq x_{th}|\pi(t)=i)$ of $x_i(t)\geq x_{th}$ is $\frac{(1-\mathbb{P}(x_{th}))p_H}{\mathbb{P}(x_{th})p_{L}+(1-\mathbb{P}(x_{th}))p_H}$.

Given $p_L\mathbb{P}(x_{th})\geq p_H(1-\mathbb{P}(x_{th}))$, the above two probabilities satisfy $\mathbf{Pr}(x_i(t)<x_{th}|\pi(t)=i)>\mathbf{Pr}(x_i(t)\geq x_{th}|\pi(t)=i)$. Then this user further compares the expected cost of choosing stochastic path $i$ and the cost of safe path 0:
\begin{align}\label{29}
    &\Scale[0.94]{\mathbb{E}_{x_i(t)}[c_i(\Bar{n}_i^{(\text{CHAR})})]-\mathbb{E}_{x_i(t)}[c_0(\Bar{n}_0^{(\text{CHAR})})]}\tag{30}\\
    =&\int_0^1(c_i(\Bar{n}_i^{(\text{CHAR})})- c_0(\Bar{n}_0^{(\text{CHAR})}))d\mathbb{P}'(\Bar{x}|\pi(t)=i)\notag\\
    =&\int_0^{x_{th}}(c_i(\Bar{n}_i^{(\text{CHAR})})- c_0(\Bar{n}_0^{(\text{CHAR})}))d\mathbb{P}'(\Bar{x}|\pi(t)=i)\notag\\&+\int_{x_{th}}^1(c_i(\Bar{n}_i^{(\text{CHAR})})- c_0(\Bar{n}_0^{(\text{CHAR})}))d\mathbb{P}'(\Bar{x}|\pi(t)=i).\notag
\end{align}
Let $\lambda_1(\Bar{x})<0$ and $\lambda_2(\Bar{x})>0$ denote the average values of two integration above. Since the cost functions $c_0(\cdot)$ and $c_i(\cdot)$ are concave with respect to $x_i(t)$, the two average values satisfy $|\lambda_1(\Bar{x})|> \lambda_2(\Bar{x})$. Taking them back to (\ref{29}), we obtain
\begin{align*}
    (\ref{29}) =&\Scale[0.94]{\lambda_1(\Bar{x})\mathbb{P}'(x_{th}|\pi(t)=i)+\lambda_2(\Bar{x})(1-\mathbb{P}'(x_{th}|\pi(t)=i))}\\=& \lambda_1(\Bar{x})\mathbf{Pr}(x_i(t)<x_{th}|i)+\lambda_2(\Bar{x})\mathbf{Pr}(x_i(t)\geq x_{th}|i)\\ \leq & \lambda_1(\Bar{x})+\lambda_2(\Bar{x}) \leq 0,
\end{align*}
where the first inequality is because of the posterior probabilities $\mathbf{Pr}(x_i(t)<x_{th}|\pi(t)=i)>\mathbf{Pr}(x_i(t)\geq x_{th}|\pi(t)=i)$ under the condition $p_L\mathbb{P}(x_{th})\geq p_H(1-\mathbb{P}(x_{th}))$, and the last inequality is due to $\lambda_1(\Bar{x})+\lambda_2(\Bar{x})<0$ given $x_i(t)\leq x_{th}$. Therefore, users given recommendation $\pi(t)=i$ believe that the expected cost on path $i$ with $\Bar{n}^{(\text{CHAR})}_i$ users is less than the cost of switching to path 0 with $\Bar{n}^{(\text{CHAR})}_0$ users there. As a result, they will not deviate from the recommendation $\pi(t)=i$. Under our assumption that $\mathbb{P}(\Bar{x})$ is mild, users may travel on any stochastic path $i$ under $\Bar{x}\sim \mathbb{P}(\Bar{x})$. Thus, there always exists $p_L,p_H\in (0,1)$ to satisfy $p_L\mathbb{P}(x_{th})\geq p_H(1-\mathbb{P}(x_{th}))$.

Similarly, if a user is recommended to choose safe path 0, his expected travel cost on path 0 is less than the cost on any stochastic path $i$, due to $\mathbf{Pr}(x_i(t)\geq x_{th}|\pi(t)=0)>\mathbf{Pr}(x_i(t)< x_{th}|\pi(t)=0)$. In summary, all users in the recommendation-only group are incentive compatible to follow the recommendations, and the expected users choosing path $i$ in this group is $(N(t)-N^{\emptyset}(t))\mathbf{Pr}(\pi(t)=i|x_i(t))$.

Regarding users in the hiding-group, they have to use their prior distribution $\mathbb{P}(\Bar{x})$ to estimate the expected costs on the two paths and make decisions as $n^{\emptyset}_i$ in (16). Therefore, the expected number of users choosing path $i$ in the hiding group is $\frac{n_i^{\emptyset}}{N(t)} N^{\emptyset}(t)$, and they are also incentive-compatible. 

\subsubsection{$\text{PoA}^{(\text{CHAR})}$ in (21)}
Recall Proposition 1 that myopic users will under-explore stochastic path $i$ if $x_i(t)\geq x_{th}$ and over-explore if $x_i(t)<x_{th}$. If $x_i(t)\geq x_{th}$ with a bad condition on stochastic path $i$, the hiding policy satisfies $n_i^{\emptyset}>n_i^*(t)$ and the recommendation probability satisfies $N(t)p_L<n_i^*(t)$, such that our CHAR can always change $n_i^{(\text{CHAR})}(t)$ to be the optimal $n_i^*(t)$ by dynamically changing the user number $N^{\emptyset}(t)$ of the hiding-group, according to the exploration number $n_i^{(\text{CHAR})}(t)$ in (18). While if $x_i(t)<x_{th}$ with a good traffic condition on path $i$, both the hiding policy and the recommendation probability satisfy $n_i^{\emptyset}>n_i^*(t)$ and $p_H N(t)>n_i^*(t)$, such that $n_i^{(\text{CHAR})}>n_i^*(t)$. In consequence, the worst-case scenario under our CHAR is users' maximum over-exploration. 

In the worst-case scenario, the expected exploration number $n_i^{(\text{CHAR})}(t)$ under our CHAR mechanism is 
\begin{align}
    n_i^{(\text{CHAR})}(t)=\frac{\Bar{N}}{M}=\frac{\Bar{N}}{M+1}+\frac{c_0(0)-c_i(0)}{M+1}, \tag{31}\label{30}
\end{align}
for any $t$. The caused immediate social cost per time slot is
\begin{align}
    c(\mathbf{n}^{(\text{CHAR})}(t))&=\sum_{i=1}^M \frac{\Bar{N}}{M} c_i\left(\frac{\Bar{N}}{M}\right)=N c_i\left(\frac{\Bar{N}}{M}\right). \tag{32}\label{31}
\end{align}
While the socially optimal policy will let some users choose path 0 to avoid the congestion on each stochastic path $i\in\{1,\cdots,M\}$. Suppose that the system has been running for a long time, such that the socially optimal policy $n^*_i(t)$ has reached its steady state. Therefore, solving $\frac{\partial c(n^*_i(t))}{\partial n_i^*(t)}=0$, we obtain the optimal expected exploration number
\begin{align}
    n^*_i(t)=\frac{\Bar{N}}{M+1}+\frac{c_0(0)-c_i(0)}{2(M+1)}\tag{33}\label{32}
\end{align}
for any stochastic path $i$ and $n^*_0(t)=\frac{\Bar{N}}{2(M+1)}$ for safe path 0.
Then we calculate the corresponding immediate social cost as:
\begin{align}
    c(\mathbf{n}^*(t))&=n^*_0(t)c_0(n^*_0(t))+\sum_{i=1}^M n^*_i(t) c_i(n^*_i(t))\tag{34}\label{33}\\
    &=\frac{\Bar{N}}{2(M+1)}(c_0(0)+c_i(0))+\frac{M}{M+1}\Bar{N}c_i(0).\notag
\end{align}
As the immediate cost under our CHAR mechanism in (\ref{31}) and that under the socially optimal policy in (\ref{33}) remain unchanged for any time slot $t$, we calculate the expected PoA as
\begin{align*}
    \text{PoA}^{(\text{CHAR})}&=\frac{c(\mathbf{n}^{(\text{CHAR})}(t))}{c(\mathbf{n}^*(t))}\\ &=1+\frac{c_0(0)-c_i(0)}{c_0(0)+c_i(0)+2Mc_i(0)}\\
    &=1+\frac{1}{2(M+1)(1+\frac{M}{\Bar{N}}V(\frac{\Bar{N}(2M+1)}{2M(M+1)}))},
\end{align*}
by inputting $c_0(0)=\frac{\Bar{N}}{M}+c_i(0)$ in (\ref{30}) and $n_i^*(t)$ in (\ref{32}).

Under the stationary distribution $\mathbb{P}(\Bar{x})$, if the travel cost of path 0 with any flow satisfies $c_0(0)\geq c_i(\frac{N_{max}}{M})$ for any stochastic path $i$, all users will opt for choosing stochastic paths, i.e., $n^{\emptyset}_i$ in (16) satisfies $n^{\emptyset}_i=\frac{\Bar{N}}{M}=n^{(m)}_i(t)$. Then any informational mechanism cannot change their expected travel cost of stochastic path $i$, and thus cannot curb their over-exploration. Therefore, $\text{PoA}^{(\text{CHAR})}$ in (21) is the minimum achievable PoA by any informational mechanism.

\subsection{Proof of Proposition 3}
We prove Proposition 3 by analyzing the same worst-case scenario with the myopic policy's zero-exploration as in Appendix E. At node $\text{D}_j$ of the linear path network in Fig. 4, a myopic user will focus on his own total cost of all the left $k+1-j$ subnetworks to make the current routing decision among paths $\{0^j,\cdots,M^j\}$. This is different from the parallel network with two nodes in Fig. 1. However, we next prove that in the worst-case scenario, all users still always choose safe path $0^j$ in any subnetwork $j\in\{0,\cdots,k\}$.

Initially, we set $\ell_{0^j}=\mathbb{E}[\ell_{i^j}(t)|x_{i^j}(t)]+V(N^j(t))$, $\alpha_L=0$ and $\alpha_H>1$ with $\mathbb{E}[\alpha_{i^j}(t)|x_{i^j}(t)]=1$ for all $j\in\{0,\cdots,k\}$. In this case, according to (8) and (9), users immediate travel cost in the $j$-th subnetwork satisfies $c_{0^j}(N^j(t))\leq c_{i^j}(0)$. Let discount factor $\rho=1$. At node $\text{D}_j$, Let $C^{j+1}(\mathbf{L}(t),\mathbf{x}(t),\mathbf{N}(t))$ denote each user's total cost-to-go from subnetwork $j+1$ to $k$. Based on the above parameter settings, we obtain
\begin{align*}
    C^{(m)}_{j+1}(\mathbf{L}(t),\mathbf{x}(t),\mathbf{N}(t))&=\min_{\{i^{j+1},\cdots,i^k\}}\sum_{l=j+1}^k \mathbb{E}[c_{i^l}(\mathbf{n}_{i^j}(t))]\\
    &=\sum_{l=j+1}^k c_{0^l}(N^j(t)),
\end{align*}
which means that the myopic policy lets all users choose path $0^{l}$ in the future subnetworks, where $l\in\{j+1,\cdots,k\}$. As all users' future cost-to-go is the same, they only need to focus on the current path with the minimum travel cost, which is path $0^j$.

On the other hand, the socially optimal policy still lets $n_{i^j}^*(t)\geq 1$ users to explore each stochastic path $i^j$ to find $\alpha_L=0$. After that, users will travel there to exploit the small travel cost.

As all myopic users always choose safe paths and the socially optimal policy lets users explore $\alpha_L=0$, we only need to focus on each subnetwork to analyze PoA. Then we can obtain the same PoA as (15) in Theorem 1. 

\subsection{Proof of Lemma 7}
At time $t$, assume that the platform has been running for a long time. Note that this is a common assumption in information-hiding mechanisms (e.g., in [18], [27], and [30]). 

If the platform hides all information, i.e., $\mathbf{L}^j(t), \mathbf{x}^j(t)$ and $N^j(t)$, from users at any node $\text{D}^j$ at any time $t$, they can only estimate the travel cost of each stochastic path $i^j$ using its steady state $\Bar{x}\sim \mathbb{P}(\Bar{x})$. At the same time, as all stochastic paths are homogeneous, the steady number of users on each path $i^j$ is the same as other stochastic paths. Therefore, each user's total cost-to-go $C^{\emptyset}_{j+1}(\bar{\mathbf{L}},\bar{\mathbf{x}},\bar{\mathbf{N}})$ from the next subnetwork $j+1$ to the last subnetwork $k$ is constant, which is the same as others and will not change with time $t$. Consequently, all users only need to focus on the current subnetwork $j$ to make routing decisions. 

As all stochastic paths are homogeneous, a user will choose randomly from these paths. Given the expected state $\bar{\mathbf{L}},\bar{\mathbf{x}}$ and $\bar{\mathbf{N}}$, the expected number of users choosing stochastic path $i^j$ is
\begin{align*}
    n^{\emptyset}_{i^j}=\frac{\bar{N}+c_{0^j}(0)-\mathbb{E}_{\Bar{x}\sim\mathbb{P}(\Bar{x})}[c_{i^j}(0)|\Bar{x}]}{M+1},
\end{align*}
which is derived by solving the Nash equilibrium as (10). However, as there are actually $N^j(t)$ users at node $\text{D}^j$, $n^{\emptyset}_{i^j}$ above should be bounded by $\frac{N^j(t)}{M}$. Finally, we obtain the expected exploration number of stochastic path $i^j$ in (22).

\subsection{Proof of Theorem 3}
Based on Lemma 7, users only focus on the current subnetwork $j$ to make routing choices. As our CHAR mechanism only recommends the path in the current subnetwork $j$ to users arriving at node $\text{D}^j$, it does not change users' cost-to-go in future subnetworks. Therefore, users are still incentive-compatible to follow the current optimal path $\pi^j(t)=i^j$ to users at any node $\text{D}^j$. Then we can follow Appendix J to prove the same PoA lower bound as Theorem 2.

\vfill

\end{document}